\documentclass[usenatbib]{mn2e}
\usepackage{epsf,graphicx,rotating}
\input{psfig}

\begin{document}
\def\la{\mathrel{\hbox{\rlap{\hbox{\lower4pt\hbox{$\sim$}}}\hbox{$<$}}}}
\def\ga{\mathrel{\hbox{\rlap{\hbox{\lower4pt\hbox{$\sim$}}}\hbox{$>$}}}}
\font\sevenrm=cmr7
\def\FeII{Fe~{\sevenrm II}}
\def\Mg{Mg~{\sevenrm II}}
\def\OII{[O~{\sevenrm II}]}
\def\OIII{[O~{\sevenrm III}]}
\def\Ha{H{\sevenrm $\alpha$}}
\def\Hb{H{\sevenrm $\beta$}}
\def\kmm{{\sevenrm KMM}}

\title[A Physical Classification Scheme for Blazars]
{A Physical Classification Scheme for Blazars}

\author[H. Landt et al.]{Hermine Landt$^{1,2,3}$, Paolo Padovani$^{1,4,5}$,
Eric S. Perlman$^{6}$, Paolo Giommi$^{7}$ \\
$^1$ Space Telescope Science Institute, 3700 San Martin Drive,
Baltimore, MD 21218, USA \\
$^2$ Hamburger Sternwarte, Gojenbergsweg 112, D-21029 Hamburg, Germany \\
$^3$ Current address: Harvard-Smithsonian Center for Astrophysics, 
60 Garden Street, Cambridge, MA  02138, USA \\
$^4$ Affiliated with the Space Telescope Division of the European Space
Agency, ESTEC, Noordwijk, The Netherlands \\
$^5$ Current address: European Southern Observatory, Karl-Schwarzschild-Str. 2,
D-85748 Garching bei M\"unchen, Germany \\ 
$^6$ Joint Center for Astrophysics, University of Maryland,
1000 Hilltop Circle, Baltimore, MD 21250, USA \\
$^7$ ASI Science Data Center, c/o ESRIN, Via G. Galilei,
I-00044 Frascati, Italy
}

\date{Accepted~~, Received~~}

\maketitle

\begin{abstract}
  
  Blazars are currently separated into BL Lacertae objects (BL Lacs)
  and flat spectrum radio quasars (FSRQ) based on the strength of
  their emission lines. This is done rather arbitrarily by defining a
  diagonal line in the Ca H\&K break value -- equivalent width plane,
  following March\~a et al. We readdress this problem and put the
  classification scheme for blazars on firm physical grounds. We study
  $\sim 100$ blazars and radio galaxies from the Deep X-ray Radio
  Blazar Survey (DXRBS) and 2 Jy radio survey and find a significant
  bimodality for the narrow emission line \OIII~$\lambda 5007$. This
  suggests the presence of two physically distinct classes of
  radio-loud AGN. We show that all radio-loud AGN, blazars and radio
  galaxies, can be effectively separated into weak- and strong-lined
  sources using the \OIII~$\lambda 5007$~--~\OII~$\lambda 3727$
  equivalent width plane. This plane allows one to disentangle
  orientation effects from intrinsic variations in radio-loud AGN.
  Based on DXRBS, the strongly beamed sources of the new class of
  weak-lined radio-loud AGN are made up of BL Lacs at the $\sim 75$
  per cent level, whereas those of the strong-lined radio-loud AGN
  include mostly ($\sim 97$ per cent) quasars.

\end{abstract}


\section{Introduction}

BL Lacertae objects (BL Lacs) and flat spectrum radio quasars (FSRQ),
commonly referred to as blazars, are the most extreme variety of
active galactic nuclei (AGN) known. Their signal properties include
irregular and rapid variability, high radio and optical polarization,
a core-dominated radio morphology, flat ($\alpha_{\rm r} \la 0.5$,
where $S_\nu \propto \nu^{-\alpha}$) radio spectra, apparent
superluminal motion, and a broad continuum extending from the radio
through the gamma-rays.

The properties of blazars can be best explained if we assume that we
view in fact a relativistic jet pointed close to our line of sight (as
originally proposed by Blandford \& Rees 1978\nocite{Bla78}). In this
scenario the large population of misdirected blazars, i.e., the
so-called `parent population', is formed by the classical double-lobed
radio galaxies.  In particular, low-luminosity radio galaxies
(Fanaroff-Riley type I [FR I]; Fanaroff \& Riley 1974\nocite{Fan74})
appear to be the parent population of BL Lacs \citep{P90, P91,
Urry91}, while high-luminosity radio galaxies (Fanaroff-Riley type II
[FR II]) are believed to be viewed mainly as FSRQ when their jets are
pointed close to our line of sight \citep{P92, Urry95}.

Blazars are currently separated into BL Lacs and FSRQ based on the
strength of their emission lines. This quantitative distinction
between blazar subclasses was prompted by observations of variable
weak emission lines in BL Lacs, which had been previously defined as
compact, radio-loud AGN with completely featureless optical spectra
\citep{Str72}. A first (and rather arbitrary) limit on the strength of
BL Lac emission lines was introduced by surveys that selected the
first complete samples of these objects, namely the 1 Jy radio survey
\citep{Sti91} and the {\it EINSTEIN} Medium Sensitivity Survey (EMSS)
at X-ray frequencies \citep{Sto91}. Both these surveys chose a value
of 5~\AA~for the maximum equivalent width of BL Lac emission lines
(applied to the rest and observer's frame respectively).

The 1 Jy survey and EMSS also introduced the first quantative
separation between blazars and radio galaxies. The 1 Jy survey used
for this purpose the radio spectral index and defined BL Lacs and
radio galaxies as sources with $\alpha_{\rm r} \le 0.5$ and
$\alpha_{\rm r} > 0.5$ respectively. The EMSS, on the other hand, did
not use radio information to classify its objects (since it searched
also for radio-quiet BL Lacs, which could not be found), and
introduced the value of the Ca H\&K break (a stellar absorption
feature) as an additional criterion. A maximum Ca H\&K break value of
25 per cent was adopted for BL Lacs in order to ensure the presence of
a substantial non-thermal jet continuum in addition to the thermal
spectrum of the elliptical host galaxy.

The Ca H\&K break limit separating blazars and radio galaxies, as well
as the equivalent width limit separating BL Lacs and FSRQ were later
revised by \citet{Marcha96}. These authors argued that any radio-loud
AGN with a Ca H\&K break below 40 per cent (and not only below 25 per
cent) should be classified as a blazar, and proposed a diagonal line
in the Ca H\&K break value -- equivalent width plane (see
Fig. \ref{marcha}) to separate BL Lacs (left of the line) and quasars
(right of the line). This diagonal line was the result of simulations
for only one object, 3C 371, which was chosen on the grounds that it
was widely accepted as a `genuine' BL Lac. Subsequently,
\citet{Sca97} showed that there was a continuity in optical continuum
and \Mg~$\lambda 2798$ luminosity between BL Lacs and FSRQ and argued
against the existence of two populations of blazars from the point of
view of emission line strengths.

The rather arbitrary separation of blazars proposed by March\~a et al.
and the results of Scarpa and Falomo make the revision of blazar
classification highly necessary. Such a revision is of particular
interest at this time since several relatively large blazar surveys
are being identified, e.g., RGB ({\it ROSAT} All Sky Survey RASS-Green
Bank; Laurent-Muehleisen et al. 1998, 1999\nocite{LM98, LM99}), REX
\citep[Radio-Emitting X-ray sources;][]{Cac99, Cac00}, the Sedentary
Survey \citep{Gio99}, CLASS (Cosmic Lens All Sky Survey) blazar sample
\citep{Marcha01, Cac02}, and DXRBS \citep[Deep X-ray Radio Blazar
Survey;][]{Per98, L01}. In this paper, we use an homogeneous sample of
blazars from DXRBS, i.e., a sample selected in a uniform and therefore
presumably unbiased way, to devise a physically based classification
scheme. DXRBS is most suitable for a physical revision of blazar
classification since it contains a large number of objects ($\sim
350$) and includes both types of blazars, BL Lacs and FSRQ. Moreover,
measurements are available on several emission line parameters as well
as on multiwavelength continuum properties.

The paper is organized as follows. In Sections 2 and 3 we present our
sample of objects and describe the emission line measurements. In
Section 4 we discuss the current blazar classification scheme proposed
by March\~a et al. and show its limitations. In Section 5 we introduce
a (new) physical classification scheme for blazars. In Sections 6 and
7 we discuss our results and present our conclusions. Throughout this
paper we have assumed cosmological parameters $H_0 = 50$ km s$^{-1}$
Mpc$^{-1}$ and $q_0 = 0$. Spectral indices have been defined as $S_\nu
\propto \nu^{-\alpha}$.

\section{The Samples}

We have selected for our analysis sources from the Deep X-ray Radio
Blazar Survey (DXRBS). DXRBS is the result of a cross-correlation of
all serendipitous X-ray sources in the publicly available {\it ROSAT}
database WGACAT (first revision: \citet{Whi95}) with a number of
publicly available radio catalogs (the 20 cm and 6 cm Green Bank
survey catalogs NORTH20CM and GB6 for the northern part of the sky,
and the 6 cm Parkes-MIT-NRAO catalog PMN for the southern part of the
sky). All sources with radio spectral index $\alpha_{\rm r} \leq 0.7$
at a few GHz and off the Galactic plane ($|b| > 10^{\circ}$) have been
selected as blazar candidates. For a detailed description of the
selection and identification procedures of the DXRBS see \citet{Per98}
and \citet{L01}. As of June 2003 the DXRBS complete sample is $\sim
95$ per cent identified. We have initially classified sources in our
sample following the scheme proposed by \citet{Marcha96} and described
in Section \ref{scheme}. The total sample (which includes also 22 'low
priority sources', i.e., sources that do not meet all selection
criteria) contains 341 objects: 261 quasars (201 flat spectrum radio
quasars [$\alpha_r \le 0.5$] and 60 steep spectrum radio quasars
[$\alpha_r > 0.5$]), 44 BL Lacs, and 36 radio galaxies.

We have selected all DXRBS BL Lacs, but have restricted the sample of
FSRQ to objects with redshifts $z \le 2.2$.  This was done in order to
compare BL Lacs and FSRQ that have the same emission lines. The
strongest emission lines common to both were \Mg~$\lambda 2798$,
\OII~$\lambda 3727$, \Hb~$\lambda 4861$, \OIII~$\lambda 5007$, and
\Ha~$\lambda 6563$. Then, $z \sim 2.2$ represents the maximum redshift
for which \Mg~can be observed in the $\sim 4000 - 9000$~\AA~range
typically covered by our optical spectra.

\subsection{The BL Lacertae Objects (BL Lacs)}

We have an optical spectrum for 28 of the 44 DXRBS BL Lacs (16 objects
were previously known sources). Additionally, we could find in the
literature information on emission lines for one more source (4C
55.17). Out of the 28 BL Lacs observed by us 5 objects do not have a
determined redshift and have been excluded from our analysis.
Therefore, the BL Lac sample under study includes 24 objects and is
listed in Table \ref{bllacs}.

\begin{table*}
\footnotesize
\begin{minipage}{180mm}
\caption{\label{bllacs} \normalsize DXRBS BL Lacs}
\renewcommand{\tabcolsep}{1.5mm}
\begin{tabular}{lcrrrrrrrrrrrrlc}
\hline
\rule{0mm}{4mm}Name & z & \multicolumn{2}{c}{MgII} &
\multicolumn{2}{c}{[OII]} & \multicolumn{2}{c}{H$\beta$} &
\multicolumn{2}{c}{[OIII]} & \multicolumn{2}{c}{H$\alpha$} &
\multicolumn{1}{c}{$L_{\rm NLR}$} & \multicolumn{1}{c}{$L_{\rm BLR}$} &
\multicolumn{1}{c}{C} & $\sigma$ \\
&& \multicolumn{1}{c}{W$_\lambda$} & \multicolumn{1}{c}{log L} &
\multicolumn{1}{c}{W$_\lambda$} & \multicolumn{1}{c}{log L} &
\multicolumn{1}{c}{W$_\lambda$} & \multicolumn{1}{c}{log L} &
\multicolumn{1}{c}{W$_\lambda$} & \multicolumn{1}{c}{log L} &
\multicolumn{1}{c}{W$_\lambda$} & \multicolumn{1}{c}{log L} & & & & \\
&& \multicolumn{1}{c}{[\AA]} & \multicolumn{1}{c}{[erg/s]} &
\multicolumn{1}{c}{[\AA]} & \multicolumn{1}{c}{[erg/s]} &
\multicolumn{1}{c}{[\AA]} & \multicolumn{1}{c}{[erg/s]} &
\multicolumn{1}{c}{[\AA]} & \multicolumn{1}{c}{[erg/s]} &
\multicolumn{1}{c}{[\AA]} & \multicolumn{1}{c}{[erg/s]} &
\multicolumn{1}{c}{[erg/s]} & \multicolumn{1}{c}{[erg/s]} & & \\[1mm]
(1) & (2) & \multicolumn{1}{c}{(3)} & \multicolumn{1}{c}{(4)} &
\multicolumn{1}{c}{(5)} & \multicolumn{1}{c}{(6)} & \multicolumn{1}{c}{(7)} &
\multicolumn{1}{c}{(8)} & \multicolumn{1}{c}{(9)} & \multicolumn{1}{c}{(10)} &
\multicolumn{1}{c}{(11)} & \multicolumn{1}{c}{(12)} & \multicolumn{1}{c}{(13)} &
\multicolumn{1}{c}{(14)} & \multicolumn{1}{c}{(15)} & \multicolumn{1}{c}{(16)} \\[0.5mm]
\hline
WGAJ0023.6$+$0417 &  0.100 &         &          &         &          & $<$ 9.8 & $<$39.56 & $<$ 7.0 & $<$39.43 & $<$16.4 & $<$39.82 & $<$40.81 & $<$40.76 & & \\
WGAJ0032.5$-$2849 &  0.324 &         &          &         &          & $<$ 1.6 & $<$41.36 &         &          &         &          &          & $<$42.76 & 0.22 & 0.08 \\
WGAJ0043.3$-$2638 &  1.002 &     6.5 &    43.50 & $<$ 0.6 & $<$42.28 &     7.5 &    43.01 &         &          &         &          & $<$43.71 &    44.61 & 0 & \\
WGAJ0100.1$-$3337 &  0.875 &    10.1 &    42.71 &         &          &         &          &         &          &         &          &          &    43.93 & & \\
WGAJ0245.2$+$1047 &  0.070 &         &          &    15.0 &    41.08 & $<$ 2.3 & $<$40.72 &    20.6 &    41.67 &    20.6 &    41.65 &    42.50 &    42.50 & 0.26 & 0.08 \\
WGAJ0313.9$+$4115 &  0.029 &         &          &         &          & $<$ 1.7 & $<$40.12 &     1.9 &    40.24 &    19.4 &    41.17 &    41.37 &    42.03 & 0.38 & 0.12 \\
WGAJ0428.8$-$3805 &  0.150 &         &          & $<$ 1.9 & $<$40.19 & $<$ 1.7 & $<$40.50 & $<$ 1.1 & $<$40.32 & $<$ 2.2 & $<$40.66 & $<$41.37 & $<$41.64 & 0.32 & 0.05 \\
WGAJ0431.9$+$1731 &  0.143 &         &          &         &          & $<$ 9.8 & $<$39.97 & $<$ 4.7 & $<$39.70 & $<$13.3 & $<$40.20 & $<$40.99 & $<$41.15 & & \\
WGAJ0528.5$-$5820 &  0.254 &         &          &         &          & $<$ 2.2 & $<$40.81 &    12.0 &    41.55 &         &          &    42.40 & $<$42.21 & 0.38 & 0.18 \\
WGAJ0533.6$-$4632 &  0.332 &         &          &         &          & $<$ 2.4 & $<$40.71 &         &          &         &          &          & $<$42.11 & 0.32 & 0.15 \\
WGAJ0558.1$+$5328 &  0.036 &         &          &         &          & $<$ 2.7 & $<$40.35 &     8.7 &    40.84 &    11.6 &    41.05 &    41.82 &    41.91 & 0.29 & 0.16 \\
WGAJ0624.7$-$3230 &  0.252 &         &          & $<$ 2.2 & $<$40.87 & $<$ 2.0 & $<$41.05 & $<$ 2.0 & $<$41.07 & $<$ 6.5 & $<$41.49 & $<$42.08 & $<$42.37 & 0.22 & 0.05 \\
WGAJ0816.0$-$0736 &  0.040 &         &          & $<$ 5.0 & $<$39.85 & $<$ 2.5 & $<$39.93 & $<$ 1.2 & $<$39.63 & $<$ 2.2 & $<$39.92 & $<$40.92 & $<$40.98 & 0.37 & 0.18 \\
WGAJ0847.2$+$1133 &  0.199 &         &          & $<$ 1.1 & $<$40.35 & $<$ 2.2 & $<$40.50 & $<$ 1.8 & $<$40.40 & $<$ 7.3 & $<$40.76 & $<$41.50 & $<$41.70 & 0 & \\
WGAJ1204.2$-$0710 &  0.185 &         &          &         &          & $<$ 4.5 & $<$41.58 &     5.1 &    41.71 &         &          &    42.54 & $<$42.98 & & \\
WGAJ1311.3$-$0521 &  0.160 &         &          & $<$ 2.7 & $<$40.50 & $<$ 1.6 & $<$40.60 & $<$ 1.2 & $<$40.47 & $<$ 1.9 & $<$40.63 & $<$41.62 & $<$41.67 & 0.33 & 0.12 \\
WGAJ1320.4$+$0140 &  1.235 &    15.2 &    43.01 &         &          &         &          &         &          &         &          &          &    44.24 & & \\
WGAJ1744.3$-$0517 &  0.310 &         &          & $<$ 3.5 & $<$40.60 & $<$ 3.3 & $<$40.72 & $<$ 2.6 & $<$40.62 &         &          & $<$41.74 & $<$42.12 & 0.20 & 0.09 \\
WGAJ1834.2$-$5948 &  0.435 &         &          & $<$13.0 & $<$40.86 & $<$12.2 & $<$40.84 & $<$ 8.0 & $<$40.68 &         &          & $<$41.94 & $<$42.24 & 0.16 & 0.36 \\
WGAJ1840.9$+$5452 &  0.646 & $<$15.3 & $<$42.83 &     5.5 &    42.38 & $<$ 8.1 & $<$42.53 &    10.9 &    42.63 &         &          &    43.61 & $<$44.00 & 0.06 & 0.12 \\
WGAJ1936.8$-$4719 &  0.264 &         &          & $<$ 0.5 & $<$40.79 & $<$ 0.9 & $<$41.04 & $<$ 0.9 & $<$40.95 &         &          & $<$41.98 & $<$42.44 & 0 & \\
WGAJ2258.3$-$5525 &  0.479 &     1.4 &    41.90 & $<$ 0.8 & $<$41.48 & $<$ 3.2 & $<$41.95 & $<$ 1.7 & $<$41.67 &         &          & $<$42.68 &    43.11 & 0 & \\
WGAJ2330.6$-$3724 &  0.279 &         &          &     0.8 &    40.18 & $<$ 1.4 & $<$40.45 &     2.4 &    40.63 &         &          &    41.52 & $<$41.85 & 0.10 & 0.04 \\
4C55.17$^a$       &  0.909 &     2.8 &    43.39 &     1.3 &    42.92 &     8.2 &    43.62 &    17.6 &    43.95 &         &          &    44.68 &    44.77 & & \\
\hline
\end{tabular}

\normalsize

Columns: (1) object name, (2) redshift, (3) and (4) \Mg~$\lambda
2798$, (5) and (6) \OII~$\lambda 3727$, (7) and (8) \Hb~$\lambda
4861$, (9) and (10) \OIII~$\lambda 5007$, (11) and (12) \Ha~$\lambda
6563$ rest frame equivalent widths and line luminosities respectively,
(13) narrow line region luminosity, (14) broad line region luminosity,
(15) Ca H\&K break value (measured in spectra $f_\lambda$ versus
$\lambda$), (16) $1\sigma$ error on Ca H\&K break value

\vspace*{0.1cm}

References for emission line data: (a) \citet{Law96}

\end{minipage}
\end{table*}

\subsection{The Flat-Spectrum Radio Quasars (FSRQ)}

We have an optical spectrum for 107 of the 178 DXRBS FSRQ with
redshifts $z \le 2.2$ (71 objects were previously known sources). Out
of these we have included in our analysis 91 objects. We have excluded
objects that: 1. had a spectrum whose wavelength range did not cover
the location of any of the emission lines analyzed here (6 objects);
2. had a spectrum covering the location of \Mg~$\lambda 2798$ only,
but either the signal-to-noise ratio (S/N) was below 3 (4 objects) or
telluric A band was present (2 objects) in that region; 3. had a
spectrum that was not taken at parallactic angle and the loss in flux
at the location of all emission lines under study was larger than 30
per cent (4 objects; see Section 3 for more details). We note that
none of these cases applied to our sample of BL Lacs. Additionally, we
could find in the literature information on emission lines for 15 of
the previously known sources. The FSRQ sample under study includes 106
objects and is listed in Table \ref{fsrq}.

\begin{table*}
\footnotesize
\begin{minipage}{180mm}
\caption{\label{fsrq} \normalsize DXRBS FSRQ with $z \le 2.2$}
\renewcommand{\tabcolsep}{1.6mm}
\begin{tabular}{lcrrrrrrrrrrrrlc}
\hline
\rule{0mm}{4mm}Name & z & \multicolumn{2}{c}{MgII} &
\multicolumn{2}{c}{[OII]} & \multicolumn{2}{c}{H$\beta$} &
\multicolumn{2}{c}{[OIII]} & \multicolumn{2}{c}{H$\alpha$} &
\multicolumn{1}{c}{$L_{\rm NLR}$} & \multicolumn{1}{c}{$L_{\rm BLR}$} &
\multicolumn{1}{c}{C} & $\sigma$ \\
&& \multicolumn{1}{c}{W$_\lambda$} & \multicolumn{1}{c}{log L} &
\multicolumn{1}{c}{W$_\lambda$} & \multicolumn{1}{c}{log L} &
\multicolumn{1}{c}{W$_\lambda$} & \multicolumn{1}{c}{log L} &
\multicolumn{1}{c}{W$_\lambda$} & \multicolumn{1}{c}{log L} &
\multicolumn{1}{c}{W$_\lambda$} & \multicolumn{1}{c}{log L} & & & & \\
&& \multicolumn{1}{c}{[\AA]} & \multicolumn{1}{c}{[erg/s]} &
\multicolumn{1}{c}{[\AA]} & \multicolumn{1}{c}{[erg/s]} &
\multicolumn{1}{c}{[\AA]} & \multicolumn{1}{c}{[erg/s]} &
\multicolumn{1}{c}{[\AA]} & \multicolumn{1}{c}{[erg/s]} &
\multicolumn{1}{c}{[\AA]} & \multicolumn{1}{c}{[erg/s]} &
\multicolumn{1}{c}{[erg/s]} & \multicolumn{1}{c}{[erg/s]} & & \\[1mm]
(1) & (2) & \multicolumn{1}{c}{(3)} & \multicolumn{1}{c}{(4)} &
\multicolumn{1}{c}{(5)} & \multicolumn{1}{c}{(6)} & \multicolumn{1}{c}{(7)} &
\multicolumn{1}{c}{(8)} & \multicolumn{1}{c}{(9)} & \multicolumn{1}{c}{(10)} &
\multicolumn{1}{c}{(11)} & \multicolumn{1}{c}{(12)} & \multicolumn{1}{c}{(13)} &
\multicolumn{1}{c}{(14)} & \multicolumn{1}{c}{(15)} & \multicolumn{1}{c}{(16)} \\[0.5mm]
\hline
WGAJ0010.5$-$3027 &  1.190 &   55.7 &  44.16 &         &          &         &          &         &          &       &         &          &    45.25 & & \\
WGAJ0012.5$-$1629 &  0.151 &        &        &         &          &    22.6 &    41.87 &    14.8 &    41.69 & 145.1 &   42.42 &    42.52 &    43.26 & 0.21 & 0.18 \\
WGAJ0029.0$+$0509 &  1.633 &   41.4 &  44.49 &         &          &         &          &         &          &       &         &          &    45.68 & & \\
WGAJ0106.7$-$1034 &  0.469 &        &        &     9.5 &    42.53 &   138.2 &    43.47 &         &          &       &         &    43.96 &    44.87 & 0 & \\
WGAJ0110.5$-$1647 &  0.781 &   34.8 &  44.20 & $<$ 3.1 & $<$42.85 &         &          &         &          &       &         & $<$44.28 &    45.41 & 0 & \\
WGAJ0126.2$-$0500 &  0.411 &   89.3 &  42.68 &    17.7 &    41.87 &    36.1 &    42.00 &    62.4 &    42.21 &       &         &    43.15 &    43.76 & 0 & \\
WGAJ0136.0$-$4044 &  0.649 &   87.3 &  42.78 &     8.5 &    41.78 &         &          &         &          &       &         &    43.21 &    43.99 & & \\
WGAJ0143.2$-$6813 &  1.223 &  101.7 &  43.56 &         &          &         &          &         &          &       &         &          &    44.71 & & \\
WGAJ0217.7$-$7347 &  1.234 &   93.1 &  43.99 &         &          &         &          &         &          &       &         &          &    45.13 & & \\
WGAJ0253.3$+$0006 &  1.339 &   53.9 &  42.65 &         &          &         &          &         &          &       &         &          &    43.86 & & \\
WGAJ0258.6$-$5052 &  0.834 &  191.9 &  42.63 &     4.9 &    41.65 & $<$21.6 & $<$42.18 &         &          &       &         &    43.09 &    43.85 & 0.09 & 0.16 \\
WGAJ0259.4$+$1926 &  0.544 &  157.4 &  43.18 &     6.5 &    41.70 &    79.7 &    42.63 &    36.3 &    42.27 &       &         &    43.11 &    44.25 & 0 & \\
WGAJ0304.9$+$0002 &  0.563 &   74.9 &  43.62 &     2.6 &    42.01 &    35.8 &    42.76 &    62.1 &    42.98 &       &         &    43.72 &    44.65 & 0 & \\
WGAJ0312.3$-$6610 &  1.384 &   50.3 &  43.37 &         &          &         &          &         &          &       &         &          &    44.53 & 0 & \\
WGAJ0314.4$-$6548 &  0.636 &  106.4 &  43.54 & $<$ 1.9 & $<$41.63 &    56.9 &    42.97 &    11.6 &    42.28 &       &         & $<$43.09 &    44.59 & 0 & \\
WGAJ0322.1$-$5205 &  0.416 &        &        &         &          &   115.8 &    43.21 &    25.4 &    42.59 & 615.1 &   43.69 &    43.33 &    44.56 & 0 & \\
WGAJ0322.2$-$5042 &  0.651 &   66.6 &  43.34 & $<$ 2.3 & $<$41.66 &    69.1 &    42.91 &    27.9 &    42.44 &       &         & $<$43.22 &    44.47 & 0 & \\
WGAJ0322.6$-$1335 &  1.468 &   93.2 &  43.44 &         &          &         &          &         &          &       &         &          &    44.61 & 0 & \\
WGAJ0357.6$-$4158 &  1.271 &  115.4 &  43.83 &         &          &         &          &         &          &       &         &          &    45.03 & & \\
WGAJ0411.0$-$1637 &  0.622 &   49.3 &  44.39 &     0.6 &    42.30 &    28.4 &    43.41 &     8.6 &    42.81 &       &         &    43.67 &    45.43 & 0 & \\
WGAJ0427.2$-$0756 &  1.375 &   88.4 &  44.07 &         &          &         &          &         &          &       &         &          &    45.77 & & \\
WGAJ0435.1$-$0811 &  0.791 &  105.5 &  43.74 &     1.7 &    41.69 &    35.7 &    42.46 &    55.3 &    42.50 &       &         &    43.27 &    44.76 & 0 & \\
WGAJ0441.8$-$4306 &  0.872 &  127.7 &  42.46 &    19.8 &    41.74 &         &          &         &          &       &         &    43.17 &    43.67 & 0 & \\
WGAJ0447.9$-$0322 &  0.774 &   52.4 &  44.85 &     1.1 &    42.97 &         &          &     4.5 &    43.05 &       &         &    44.13 &    46.06 & 0 & \\
WGAJ0448.6$-$2203 &  0.496 &  127.0 &  42.03 &    39.4 &    41.39 &         &          &         &          &       &         &    42.82 &    43.24 & 0 & \\
WGAJ0510.0$+$1800 &  0.416 &   55.8 &  42.95 &     5.7 &    42.18 &         &          &         &          &       &         &    43.61 &    44.09 & 0.02 & 0.05 \\
WGAJ0535.1$-$0239 &  1.033 &   83.1 &  44.30 &         &          &         &          &         &          &       &         &          &    45.58 & 0 & \\
WGAJ0539.0$-$3427 &  0.263 &        &        &         &          &   263.7 &    41.91 &         &          & 169.4 &   41.82 &          &    42.92 & 0.40 & 0.22 \\
WGAJ0546.6$-$6415 &  0.323 &        &        &     0.8 &    42.08 &    93.0 &    43.84 &    24.9 &    43.25 & 442.2 &   44.28 &    43.96 &    45.15 & 0 & \\
WGAJ0600.5$-$3937 &  1.661 &   20.7 &  44.25 &         &          &         &          &         &          &       &         &          &    45.62 & & \\
WGAJ0631.9$-$5404 &  0.193 &        &        &     4.2 &    41.82 &    77.1 &    42.96 &    33.5 &    42.59 & 378.0 &   43.50 &    43.37 &    44.34 & 0 & \\
WGAJ0648.2$-$4347 &  1.029 &   56.7 &  44.45 &         &          &         &          &         &          &       &         &          &    45.61 & 0 & \\
WGAJ0724.3$-$0715 &  0.271 &        &        & $<$ 4.6 & $<$42.05 &         &          &    10.2 &    42.29 &  38.6 &   42.57 & $<$43.28 &    43.43 & 0.01 & 0.15 \\
WGAJ0744.8$+$2920 &  1.168 &   15.2 &  44.36 &         &          &         &          &         &          &       &         &          &    45.57 & 0 & \\
WGAJ0747.0$-$6744 &  1.025 &   85.9 &  43.47 &         &          &         &          &         &          &       &         &          &    44.68 & & \\
WGAJ0748.2$-$5257 &  1.802 &   37.8 &  44.69 &         &          &         &          &         &          &       &         &          &    45.91 & & \\
WGAJ0751.0$-$6726 &  1.237 &   42.9 &  44.33 & $<$ 1.6 & $<$42.64 &         &          &         &          &       &         & $<$44.07 &    45.57 & & \\
WGAJ0927.7$-$0900 &  0.254 &        &        &         &          &    44.7 &    42.05 &    24.7 &    41.77 & 173.1 &   42.37 &    42.59 &    43.29 & & \\
WGAJ0937.2$+$5008 &  0.275 &        &        & $<$ 1.2 & $<$41.29 &     7.8 &    42.05 &     2.4 &    41.51 &  30.6 &   42.25 & $<$42.51 &    43.25 & 0 & \\
WGAJ0954.4$-$0503 &  0.660 &   31.9 &  41.39 &    10.8 &    41.20 & $<$16.3 & $<$41.70 &    10.8 &    41.51 &       &         &    42.46 &    42.61 & 0.32 & 0.15 \\
WGAJ1003.9$+$3244 &  1.682 &   33.6 &  44.23 &         &          &         &          &         &          &       &         &          &    45.68 & & \\
WGAJ1006.5$+$0509 &  1.216 &  177.8 &  43.32 &         &          &         &          &         &          &       &         &          &    44.67 & & \\
WGAJ1010.8$-$0201 &  0.896 &   64.4 &  43.88 &         &          &    76.0 &    43.45 &    23.2 &    42.90 &       &         &    43.73 &    45.12 & & \\
WGAJ1011.5$-$0423 &  1.588 &   32.8 &  44.40 &         &          &         &          &         &          &       &         &          &    45.68 & & \\
WGAJ1025.9$+$1253 &  0.663 &   70.4 &  43.52 &     2.4 &    42.01 &    22.9 &    42.57 &    23.5 &    42.52 &       &         &    43.38 &    44.51 & 0 & \\
WGAJ1026.4$+$6746 &  1.181 &  124.7 &  45.12 & $<$ 5.0 & $<$43.33 &         &          &         &          &       &         & $<$44.76 &    46.33 & 0 & \\
WGAJ1028.5$-$0236 &  0.476 &   14.9 &  42.38 &         &          & $<$ 4.3 & $<$41.83 &     4.1 &    41.79 &       &         &    42.60 &    43.59 & 0.03 & 0.02 \\
WGAJ1032.1$-$1400 &  1.039 &   63.8 &  43.86 &         &          &         &          &         &          &       &         &          &    45.07 & & \\
WGAJ1035.0$+$5652 &  0.577 &   90.0 &  42.88 & $<$ 4.2 & $<$41.51 & $<$32.0 & $<$42.10 & $<$ 8.0 & $<$41.53 &       &         & $<$42.65 &    44.09 & 0 & \\
WGAJ1101.8$+$6241 &  0.663 &        &        &     7.8 &    42.97 &    43.3 &    43.30 &    40.3 &    43.25 &       &         &    44.22 &    44.72 & 0 & \\
WGAJ1104.8$+$6038 &  1.373 &   72.1 &  44.61 & $<$ 6.1 & $<$43.05 &         &          &         &          &       &         & $<$44.48 &    45.82 & & \\
WGAJ1105.3$-$1813 &  0.578 &   50.7 &  43.40 & $<$ 2.5 & $<$41.79 &    31.7 &    42.61 &    17.7 &    42.32 &       &         & $<$43.17 &    44.42 & 0 & \\
WGAJ1112.5$-$3745 &  0.979 &   92.0 &  44.24 &     3.0 &    42.57 &         &          &         &          &       &         &    44.00 &    45.35 & 0 & \\
WGAJ1206.2$+$2823 &  0.708 &   62.1 &  43.44 & $<$ 3.1 & $<$41.98 &         &          &         &          &       &         & $<$43.41 &    44.65 & 0 & \\
WGAJ1213.2$+$1443 &  0.714 &  179.7 &  43.69 & $<$ 6.1 & $<$42.05 &         &          &    66.5 &    42.75 &       &         & $<$43.55 &    44.90 & 0 & \\
WGAJ1217.1$+$2925 &  0.974 &   62.8 &  43.32 &    44.6 &    42.52 &         &          &         &          &       &         &    43.95 &    44.54 & 0 & \\
WGAJ1223.9$+$0650 &  1.189 &   43.9 &  43.59 &         &          &         &          &         &          &       &         &          &    44.74 & & \\
\hline
\end{tabular}
\end{minipage}
\end{table*}

\begin{table*}
\footnotesize
\begin{minipage}{180mm}
\renewcommand{\tabcolsep}{1.6mm}
\setcounter{table}{2}
\contcaption{}
\begin{tabular}{lcrrrrrrrrrrrrlc}
\hline
\rule{0mm}{4mm}Name & z & \multicolumn{2}{c}{MgII} &
\multicolumn{2}{c}{[OII]} & \multicolumn{2}{c}{H$\beta$} &
\multicolumn{2}{c}{[OIII]} & \multicolumn{2}{c}{H$\alpha$} &
\multicolumn{1}{c}{$L_{\rm NLR}$} & \multicolumn{1}{c}{$L_{\rm BLR}$} &
\multicolumn{1}{c}{C} & $\sigma$ \\
&& \multicolumn{1}{c}{W$_\lambda$} & \multicolumn{1}{c}{log L} &
\multicolumn{1}{c}{W$_\lambda$} & \multicolumn{1}{c}{log L} &
\multicolumn{1}{c}{W$_\lambda$} & \multicolumn{1}{c}{log L} &
\multicolumn{1}{c}{W$_\lambda$} & \multicolumn{1}{c}{log L} &
\multicolumn{1}{c}{W$_\lambda$} & \multicolumn{1}{c}{log L} & & & & \\
&& \multicolumn{1}{c}{[\AA]} & \multicolumn{1}{c}{[erg/s]} &
\multicolumn{1}{c}{[\AA]} & \multicolumn{1}{c}{[erg/s]} &
\multicolumn{1}{c}{[\AA]} & \multicolumn{1}{c}{[erg/s]} &
\multicolumn{1}{c}{[\AA]} & \multicolumn{1}{c}{[erg/s]} &
\multicolumn{1}{c}{[\AA]} & \multicolumn{1}{c}{[erg/s]} &
\multicolumn{1}{c}{[erg/s]} & \multicolumn{1}{c}{[erg/s]} & & \\[1mm]
(1) & (2) & \multicolumn{1}{c}{(3)} & \multicolumn{1}{c}{(4)} &
\multicolumn{1}{c}{(5)} & \multicolumn{1}{c}{(6)} & \multicolumn{1}{c}{(7)} &
\multicolumn{1}{c}{(8)} & \multicolumn{1}{c}{(9)} & \multicolumn{1}{c}{(10)} &
\multicolumn{1}{c}{(11)} & \multicolumn{1}{c}{(12)} & \multicolumn{1}{c}{(13)} &
\multicolumn{1}{c}{(14)} & \multicolumn{1}{c}{(15)} & \multicolumn{1}{c}{(16)} \\[0.5mm]
\hline
WGAJ1300.7$-$3253 &  1.256 &   61.2 &  44.12 & $<$ 4.1 & $<$42.70 &         &          &         &          &       &         & $<$44.13 &    45.33 & & \\
WGAJ1306.6$-$2428 &  0.666 &  109.8 &  42.52 &    10.2 &    41.43 &    63.0 &    42.15 &    54.0 &    42.07 &       &         &    42.89 &    43.62 & 0.03 & 0.13 \\
WGAJ1314.0$-$3304 &  0.484 &        &$>$43.34$^\ast$&    26.3 &    42.35 & $<$12.9 & $<$42.03 &   103.8 &    42.89 &       &         &    43.74 & $>$44.55$^\ast$ & 0.05 & 0.15 \\
WGAJ1315.1$+$2841 &  1.576 &   24.5 &  44.35 &         &          &         &          &         &          &       &         &          &    45.56 & & \\
WGAJ1324.0$-$3623 &  0.739 &   47.2 &  44.18 &     2.3 &    42.72 &    76.5 &    43.93 &    27.0 &    43.47 &       &         &    44.25 &    45.37 & 0 & \\
WGAJ1359.6$+$4010 &  0.407 &        &        &     4.3 &    41.79 &    11.4 &    42.34 &    12.8 &    42.39 & 150.0 &   43.22 &    43.22 &    44.02 & 0.13 & 0.16 \\
WGAJ1416.4$+$1242 &  0.335 &        &        & $<$ 1.9 & $<$41.63 &    67.4 &    42.83 &    21.0 &    42.32 &       &         & $<$43.12 &    44.23 & 0 & \\
WGAJ1442.3$+$5236 &  1.800 &  222.9 &  44.67 &         &          &         &          &         &          &       &         &          &    45.90 & & \\
WGAJ1506.6$-$4008 &  1.031 &   45.8 &  44.25 &         &          &         &          &         &          &       &         &          &    45.53 & 0 & \\
WGAJ1509.5$-$4340 &  0.776 &   40.5 &  43.94 & $<$ 1.6 & $<$42.30 &    85.6 &    43.67 &    20.8 &    43.06 &       &         & $<$43.84 &    45.12 & 0 & \\
WGAJ1525.3$+$4201 &  1.189 &   73.6 &  44.30 & $<$ 2.5 & $<$42.78 &         &          &         &          &       &         & $<$44.21 &    45.51 & & \\
WGAJ1543.6$+$1847 &  1.396 &   45.1 &  44.13 &         &          &         &          &         &          &       &         &          &    45.38 & & \\
WGAJ1606.0$+$2031 &  0.383 &        &        &    25.3 &    41.64 & $<$18.6 & $<$41.70 &    15.2 &    41.71 & 117.9 &   42.56 &    42.80 &    43.42 & 0.24 & 0.24 \\
WGAJ1610.3$-$3958 &  0.518 &   25.0 &  43.04 & $<$ 1.3 & $<$41.72 &    18.5 &    42.84 &         &          &       &         & $<$43.15 &    44.25 & 0 & \\
WGAJ1629.7$+$2117 &  0.833 &  103.7 &  43.20 &    28.9 &    42.40 &         &          &         &          &       &         &    43.83 &    44.41 & 0 & \\
WGAJ1656.6$+$5321 &  1.555 &   54.0 &  44.25 &         &          &         &          &         &          &       &         &          &    45.47 & & \\
WGAJ1656.6$+$6012 &  0.623 &  100.4 &  43.76 &     4.9 &    42.23 &    35.1 &    42.98 &    17.3 &    42.67 &       &         &    43.56 &    44.79 & 0 & \\
WGAJ1804.7$+$1755 &  0.435 &   73.2 &  43.73 & $<$ 3.6 & $<$42.24 &    44.6 &    43.09 &    46.7 &    43.09 &       &         & $<$43.85 &    44.75 & 0 & \\
WGAJ1808.2$-$5011 &  1.606 &    8.1 &  43.55 &         &          &         &          &         &          &       &         &          &    45.43 & & \\
WGAJ1826.1$-$3650 &  0.888 &   28.1 &  43.48 &         &          &         &          &         &          &       &         &          &    44.69 & & \\
WGAJ1827.1$-$4533 &  1.244 &   64.2 &  44.60 &         &          &         &          &         &          &       &         &          &    45.77 & & \\
WGAJ1911.8$-$2102 &  1.420 &   14.5 &  43.80 &         &          &         &          &         &          &       &         &          &    45.22 & & \\
WGAJ1938.4$-$4657 &  0.805 &   48.8 &  41.70 &    14.4 &    41.34 &         &          &         &          &       &         &    42.77 &    42.91 & 0.28 & 0.15 \\
WGAJ2109.7$-$1332 &  1.226 &   69.6 &  44.75 & $<$ 5.0 & $<$43.41 &         &          &         &          &       &         & $<$44.84 &    45.96 & & \\
WGAJ2154.1$-$1502 &  1.208 &   39.4 &  44.87 & $<$ 1.4 & $<$43.17 &         &          &         &          &       &         & $<$44.60 &    46.28 & & \\
WGAJ2157.7$+$0650 &  0.625 &        &        &    39.4 &    42.18 &   132.9 &    42.69 &   187.1 &    42.83 &       &         &    43.65 &    44.09 & 0 & \\
WGAJ2304.8$-$3624 &  0.962 &  139.1 &  43.12 &    23.5 &    42.36 &         &          &         &          &       &         &    43.79 &    44.33 & 0 & \\
WGAJ2320.6$+$0032 &  1.894 &   61.9 &  43.52 &         &          &         &          &         &          &       &         &          &    45.49 & & \\
WGAJ2322.0$+$2113 &  0.707 &   45.1 &  44.18 & $<$ 1.6 & $<$42.55 &         &          &    12.3 &    42.73 &       &         & $<$43.75 &    45.40 & & \\
WGAJ2329.0$+$0834 &  0.948 &   70.3 &  43.05 &         &          &         &          &         &          &       &         &          &    44.26 & & \\
WGAJ2333.2$-$0131 &  1.062 &   90.7 &  43.86 &         &          &         &          &         &          &       &         &          &    44.94 & & \\
WGAJ2349.9$-$2552 &  0.844 &  108.5 &  43.74 &     6.0 &    42.34 &         &          &         &          &       &         &    43.78 &    44.95 & 0 & \\
WGAJ2354.2$-$0957 &  0.989 &   57.8 &  42.76 &    12.1 &    42.02 &         &          &         &          &       &         &    43.45 &    43.97 & 0 & \\
1Jy 0112$-$017$^a$&  1.381 &   21.0 &  43.68 &         &          &         &          &         &          &       &         &          &    44.98 & & \\
1Jy 0119$+$041$^{b,c}$&0.637&  36.0 &  43.82 &         &          &         &    43.25 &         &    42.80 &       &         &    43.63 &    44.87 & & \\
1Jy 0514$-$459$^d$&  0.194 &        &        &         &          &     6.7 &    42.33 &     5.0 &    42.18 &  35.2 &   43.00 &    42.95 &    43.83 & & \\
1Jy 0850$+$581$^e$&  1.322 &   60.1 &  45.20 &     1.1 &    43.15 &         &          &         &          &       &         &    44.58 &    46.34 & & \\
1Jy 0859$+$470$^e$&  1.462 &   73.9 &  44.84 &     3.3 &    43.20 &         &          &         &          &       &         &    44.64 &    46.07 & & \\
1Jy 1637$+$574$^e$&  0.750 &   19.9 &  44.39 &     1.8 &    43.04 &         &    44.57 &    11.6 &    43.72 &       &         &    44.52 &    45.79 & & \\
1Jy 1638$+$398$^f$&  1.666 &    2.3 &  43.39 &         &          &         &          &         &          &       &         &          &    44.65 & & \\
1Jy 1725$+$044$^g$&  0.293 &        &        &         &          &         &    43.23 &         &    42.15 &       &   43.82 &    42.93 &    44.67 & & \\
1Jy 2344$+$092$^a$&  0.673 &   41.2 &  44.62 &         &    42.92 &         &    44.45 &         &          &       &         &    44.36 &    45.99 & & \\
S5 0743$+$74$^d$  &  1.629 &   12.6 &  44.19 &         &          &         &          &         &          &       &         &          &    45.40 & & \\
S5 1027$+$74$^h$  &  0.123 &        &        &     3.6 &    40.46 &    39.2 &    41.91 &    49.9 &    42.01 &       &   42.37 &    42.69 &    43.20 & & \\
3C 345$^e$        &  0.594 &   55.6 &  44.86 &     0.6 &    42.58 &         &    43.23 &     5.4 &    43.48 &       &         &    44.23 &    45.95 & & \\
4C 38.41$^e$      &  1.814 &   29.3 &  45.15 &         &          &         &          &         &          &       &         &          &    46.67 & & \\
PKS 2059$+$034$^a$&  1.013 &   34.3 &  44.26 &         &          &         &          &         &          &       &         &          &    45.49 & & \\
OY$-$106$^f$      &  0.618 &   17.9 &  43.19 &     9.3 &    42.84 &         &          &    38.3 &    43.39 &       &         &    44.24 &    44.40 & & \\
\hline
\end{tabular}

\normalsize

Columns: (1) object name, (2) redshift, (3) and (4) \Mg~$\lambda
2798$, (5) and (6) \OII~$\lambda 3727$, (7) and (8) \Hb~$\lambda
4861$, (9) and (10) \OIII~$\lambda 5007$, (11) and (12) \Ha~$\lambda
6563$ rest frame equivalent widths and line luminosities respectively,
(13) narrow line region luminosity, (14) broad line region luminosity,
(15) Ca H\&K break value (measured in spectra $f_\lambda$ versus
$\lambda$), (16) $1\sigma$ error on Ca H\&K break value

\vspace*{0.1cm}

References for emission line data: (a) \citet{Bal89}, (b)
\citet{Rich80}, (c) \citet{Jack91}, (d) \citet{Sti93a}, (e)
\citet{Law96}, (f) \citet{Sti89}, (g) \citet{Jack95}, (h)
\citet{Sti93b}

\vspace*{0.1cm}

$\ast$ lower limit since flux loss due to atmospheric differential
refraction $>30$ per cent

\end{minipage}
\end{table*}

\subsection{The Steep-Spectrum Radio Quasars (SSRQ)} \label{ssrqsec}

For our studies in Section \ref{bimodal} we have additionally selected
DXRBS SSRQ with redshifts $z \le 1.2$, the maximum redshift at which
the Ca H\&K break is observable in our spectra (40 objects). We have
an optical spectrum for 32 sources (8 objects were previously known)
and included all of these in our analysis. Additionally, we could find
in the literature information on emission lines for two more sources.
The SSRQ sample under study includes 34 objects and is listed in Table
\ref{ssrq}.

\begin{table*}
\footnotesize
\begin{minipage}{180mm}
\caption{\label{ssrq} \normalsize DXRBS SSRQ with $z\le1.2$}
\renewcommand{\tabcolsep}{1.6mm}
\begin{tabular}{lcrrrrrrrrrrrrlc}
\hline
\rule{0mm}{4mm}Name & z & \multicolumn{2}{c}{MgII} &
\multicolumn{2}{c}{[OII]} & \multicolumn{2}{c}{H$\beta$} &
\multicolumn{2}{c}{[OIII]} & \multicolumn{2}{c}{H$\alpha$} &
\multicolumn{1}{c}{$L_{\rm NLR}$} & \multicolumn{1}{c}{$L_{\rm BLR}$} &
\multicolumn{1}{c}{C} & $\sigma$ \\
&& \multicolumn{1}{c}{W$_\lambda$} & \multicolumn{1}{c}{log L} &
\multicolumn{1}{c}{W$_\lambda$} & \multicolumn{1}{c}{log L} &
\multicolumn{1}{c}{W$_\lambda$} & \multicolumn{1}{c}{log L} &
\multicolumn{1}{c}{W$_\lambda$} & \multicolumn{1}{c}{log L} &
\multicolumn{1}{c}{W$_\lambda$} & \multicolumn{1}{c}{log L} & & & & \\
&& \multicolumn{1}{c}{[\AA]} & \multicolumn{1}{c}{[erg/s]} &
\multicolumn{1}{c}{[\AA]} & \multicolumn{1}{c}{[erg/s]} &
\multicolumn{1}{c}{[\AA]} & \multicolumn{1}{c}{[erg/s]} &
\multicolumn{1}{c}{[\AA]} & \multicolumn{1}{c}{[erg/s]} &
\multicolumn{1}{c}{[\AA]} & \multicolumn{1}{c}{[erg/s]} &
\multicolumn{1}{c}{[erg/s]} & \multicolumn{1}{c}{[erg/s]} & & \\[1mm]
(1) & (2) & \multicolumn{1}{c}{(3)} & \multicolumn{1}{c}{(4)} &
\multicolumn{1}{c}{(5)} & \multicolumn{1}{c}{(6)} & \multicolumn{1}{c}{(7)} &
\multicolumn{1}{c}{(8)} & \multicolumn{1}{c}{(9)} & \multicolumn{1}{c}{(10)} &
\multicolumn{1}{c}{(11)} & \multicolumn{1}{c}{(12)} & \multicolumn{1}{c}{(13)} &
\multicolumn{1}{c}{(14)} & \multicolumn{1}{c}{(15)} & \multicolumn{1}{c}{(16)} \\[0.5mm]
\hline
WGAJ0034.4$-$2133 & 0.764 & 108.8 &   42.66 &     20.4 &    42.06 &          &          &       &       &       &       &    43.50 &    43.88 & 0.24 & 0.18 \\
WGAJ0211.9$-$7351 & 0.789 &  48.6 &   43.33 & $<$  1.5 & $<$41.69 &          &          &       &       &       &       & $<$43.12 &    44.54 & 0    &      \\
WGAJ0251.9$-$2051 & 0.761 &       &$>$43.79$^\ast$ &      0.7 &    41.74 &          &          &       &       &       &       &    43.17 & $>$45.01$^\ast$ & 0    &      \\
WGAJ0307.7$-$4717 & 0.599 & 120.1 &   43.28 &      6.9 &    41.93 &          &          &       &       &       &       &    43.36 &    44.49 & 0.05 & 0.10 \\
WGAJ0321.6$-$6641 & 0.546 &  54.3 &   43.27 &      9.1 &    42.24 &    119.0 &    43.03 & 143.6 & 43.05 &       &       &    43.82 &    44.46 & 0    &      \\
WGAJ0325.0$-$4927 & 0.259 &       &         &          &          &     43.7 &    42.47 &       &       & 199.4 & 42.99 &          &    43.84 & 0    &      \\
WGAJ0355.6$-$1026 & 0.965 & 186.8 &   43.42 &     13.2 &    42.15 & $<$ 31.8 & $<$42.34 &  97.2 & 42.71 &       &       &    43.55 &    44.53 & 0    &      \\
WGAJ0414.0$-$1224 & 0.569 &  51.0 &   43.46 &          &          &    106.4 &    43.30 &  39.5 & 42.85 &       &       &    43.68 &    44.69 & 0    &      \\
WGAJ0414.0$-$1307 & 0.463 &       &         &     67.7 &    42.10 &          &          &       &       &       &       &    43.53 &    43.60 & 0.38 & 0.21 \\
WGAJ0518.2$+$0624 & 0.891 &  57.6 &   44.28 &          &          &          &          &       &       &       &       &          &    45.49 & & \\
WGAJ0533.7$-$5817 & 0.757 &  87.1 &   44.02 &      4.0 &    42.43 &          &          &       &       &       &       &    43.86 &    45.23 & 0    &      \\
WGAJ0646.8$+$6807 & 0.927 &  79.9 &   44.39 &      8.8 &    43.05 &          &          &       &       &       &       &    44.49 &    45.60 & 0    &      \\
WGAJ0829.5$+$0858 & 0.866 & 120.0 &   43.07 &          &          &          &          &       &       &       &       &          &    44.28 & 0.13 & 0.18 \\
WGAJ0900.2$-$2817 & 0.894 &  62.3 &   44.20 &          &          &          &          &       &       &       &       &          &    45.42 & & \\
WGAJ0908.2$+$5031 & 0.917 & 105.4 &   43.22 &          &          &          &          &       &       &       &       &          &    44.43 & & \\
WGAJ0931.9$+$5533 & 0.266 &       &         &      2.4 &    41.96 &     58.5 &    43.08 &  11.1 & 42.33 &       &       &    43.25 &    44.43 & 0    &      \\
WGAJ1006.1$+$3236 & 1.020 &  62.4 &   43.98 &          &          &          &          &       &       &       &       &          &    45.19 & 0    &      \\
WGAJ1222.6$+$2934 & 0.787 &  71.1 &   43.92 &     12.3 &    42.90 &     16.8 &    42.62 &  33.6 & 42.89 &       &       &    44.02 &    44.94 & 0    &      \\
WGAJ1225.5$+$0715 & 1.120 &  94.8 &   43.39 &     12.3 &    42.48 &          &          &       &       &       &       &    43.91 &    44.42 & 0    &      \\
WGAJ1332.7$+$4722 & 0.668 &  41.4 &   43.51 &      4.2 &    42.35 &     40.2 &    43.10 &  12.0 & 42.57 &       &       &    43.57 &    44.65 & 0    &      \\
WGAJ1353.2$-$4720 & 0.550 &  92.9 &   43.79 &      9.7 &    42.56 &     38.1 &    42.93 &  63.9 & 43.12 &       &       &    43.97 &    44.76 & 0    &      \\
WGAJ1404.2$+$3413 & 0.937 &  69.2 &   44.74 &          &          &          &          &       &       &       &       &          &    45.95 & 0    &      \\
WGAJ1420.6$+$0650 & 0.236 &       &         &     33.2 &    41.69 & $<$  5.7 & $<$41.33 &  11.3 & 41.63 &  43.7 & 42.16 &    42.80 &    43.02 & 0.33 & 0.14 \\
WGAJ1423.3$+$4830 & 0.569 & 805.6 &   43.77 &     36.3 &    42.47 & $<$ 21.2 & $<$42.37 &  98.8 & 42.95 &       &       &    43.82 &    44.98 & 0    &      \\
WGAJ1427.9$+$3247 & 0.568 &       &         &      7.0 &    43.11 &     47.8 &    43.51 &  94.4 & 43.75 &       &       &    44.56 &    44.91 & 0.07 & 0.33 \\
WGAJ1626.6$+$5809 & 0.748 &  33.2 &   44.44 & $<$  1.0 & $<$42.65 &     74.9 &    44.24 &  28.0 & 43.82 &       &       & $<$44.53 &    45.65 & 0    &      \\
WGAJ1648.4$+$4104 & 0.851 &  44.8 &   43.78 &      9.7 &    42.98 &    108.1 &    43.74 & 110.2 & 43.75 &       &       &    44.53 &    45.01 & 0    &      \\
WGAJ1722.3$+$3103 & 0.305 &       &         &    111.9 &    41.75 &     33.0 &    41.92 &  20.7 & 41.72 & 432.2 & 43.07 &    42.87 &    43.80 & 0.32 & 0.51 \\
WGAJ2056.4$-$5819 & 1.139 &  49.1 &   44.61 &          &          &          &          &       &       &       &       &          &    45.68 & & \\
WGAJ2201.6$-$5646 & 0.410 & 120.6 &   43.21 &          &          &     41.8 &    42.47 &  19.1 & 42.10 & 565.2 & 43.40 &    42.88 &    44.27 & 0    &      \\
WGAJ2239.7$-$0631 & 0.264 &       &         &      9.5 &    41.30 &     14.2 &    41.65 &  24.5 & 41.87 & 159.0 & 42.67 &    42.71 &    43.44 & 0.14 & 0.11 \\
WGAJ2347.6$+$0852 & 0.292 &       &         &      4.6 &    42.03 &     43.3 &    42.98 &  51.1 & 43.06 & 281.0 & 43.22 &    43.78 &    44.19 & 0    &      \\
PKS 2058$-$425$^a$& 0.221 &       &         &          &          &          &          &  13.9 &       &       &       &          &          & & \\
PKS 2352$-$342$^a$& 0.702 &  38.8 &         &          &          &          &          &       &       &       &       &          &          & & \\
\hline
\end{tabular}

\normalsize

Columns: (1) object name, (2) redshift, (3) and (4) \Mg~$\lambda
2798$, (5) and (6) \OII~$\lambda 3727$, (7) and (8) \Hb~$\lambda
4861$, (9) and (10) \OIII~$\lambda 5007$, (11) and (12) \Ha~$\lambda
6563$ rest frame equivalent widths and line luminosities respectively,
(13) narrow line region luminosity, (14) broad line region luminosity,
(15) Ca H\&K break value (measured in spectra $f_\lambda$ versus
$\lambda$), (16) $1\sigma$ error on Ca H\&K break value

\vspace*{0.1cm}

References for emission line data: (a) \citet{Wil86}

\vspace*{0.1cm}

$\ast$ lower limit since flux loss due to atmospheric differential
refraction $>30$ per cent

\end{minipage}
\end{table*}

\subsection{The Radio Galaxies} \label{rgsec}

For our studies in Section \ref{bimodal} we have additionally selected
DXRBS radio galaxies (36 objects). We have an optical spectrum for 18
sources (18 objects were previously known), but have excluded from our
analysis 8 objects with no emission lines and a spectrum with a
resolution too low to allow for a derivation of non-detection limits
(see Section \ref{lines} for more details). The sample of radio
galaxies under study includes 10 objects and is listed in Table
\ref{nlrg}.

\begin{center}
\begin{table*}
\footnotesize
\begin{minipage}{120mm}
\caption{\label{nlrg} \normalsize DXRBS Radio Galaxies}
\renewcommand{\tabcolsep}{2mm}
\begin{tabular}{lcrrrrrcc}
\hline
\rule{0mm}{4mm}Name & z & \multicolumn{2}{c}{[OII]} & \multicolumn{2}{c}{[OIII]} &
\multicolumn{1}{c}{$L_{\rm NLR}$} & C & $\sigma$ \\
&& \multicolumn{1}{c}{W$_\lambda$} & \multicolumn{1}{c}{log L} &
\multicolumn{1}{c}{W$_\lambda$} & \multicolumn{1}{c}{log L} &&& \\
&& \multicolumn{1}{c}{[\AA]} & \multicolumn{1}{c}{[erg/s]} &
\multicolumn{1}{c}{[\AA]} & \multicolumn{1}{c}{[erg/s]} &
\multicolumn{1}{c}{[erg/s]} && \\[1mm]
(1) & (2) & \multicolumn{1}{c}{(3)} & \multicolumn{1}{c}{(4)} & \multicolumn{1}{c}{(5)} &
\multicolumn{1}{c}{(6)} & \multicolumn{1}{c}{(7)} & (8) & (9) \\[0.5mm]
\hline
WGAJ0204.8$+$1514 &  0.833 &    49.1 &   42.35 &    85.7 &   42.93 &   43.77 & 0.21 & 0.37 \\
WGAJ0247.9$+$1845 &  0.301 & $<$ 5.8 &$<$40.71 & $<$ 1.8 &$<$40.63 &$<$41.82 & 0.47 & 0.26 \\
WGAJ0500.0$-$3040 &  0.417 &    20.5 &   42.12 &    57.2 &   42.77 &   43.58 & 0.15 & 0.08 \\
WGAJ0605.8$-$7556 &  0.458 &    39.8 &   41.21 &    32.9 &   41.18 &   42.33 & 0.03 & 0.24 \\
WGAJ1120.4$+$5855 &  0.158 &         &         &    43.2 &   42.71 &   43.44 &      &      \\
WGAJ1229.4$+$2711 &  0.490 &    74.5 &   42.69 &    53.0 &   42.57 &   43.78 & 0.42 & 0.53 \\
WGAJ1835.5$-$6539 &  0.554 & $<$ 4.8 &$<$40.46 &    42.5 &   41.53 &   42.36 & 0.11 & 0.23 \\
WGAJ2131.9$-$0556 &  0.085 & $<$ 9.2 &$<$39.64 & $<$ 2.3 &$<$39.53 &$<$40.74 & 0.43 & 0.20 \\
WGAJ2205.2$-$0004 &  0.827 &    41.1 &   42.15 &         &         &   43.59 & 0.36 & 0.75 \\
WGAJ2303.5$-$5126 &  0.426 &    11.9 &   41.70 &    18.9 &   42.06 &   42.99 & 0.12 & 0.19 \\
\hline
\end{tabular}

\normalsize

Columns: (1) object name, (2) redshift, (3) and (4) \OII~$\lambda
3727$, (5) and (6) \OIII~$\lambda 5007$ rest frame equivalent widths
and line luminosities respectively, (7) narrow line region luminosity,
(8) Ca H\&K break value (measured in spectra $f_\lambda$ versus
$\lambda$), (9) $1\sigma$ error on Ca H\&K break value

\end{minipage}
\end{table*}
\end{center}

\section{Emission Line Measurements} \label{lines}

Tables 1 to 4 list the rest frame equivalent widths and line
luminosities for \Mg~$\lambda 2798$, \OII~$\lambda 3727$, \Hb~$\lambda
4861$, \OIII~$\lambda 5007$, and \Ha~$\lambda 6563$ for our sample of
DXRBS sources.

In the individual spectra a cubic spline was first fitted
interactively to the continuum over the full wavelength range covered
using the IRAF task {\it noao.onedspec.sfit}. The equivalent widths
($W_{\lambda}$) and line fluxes were then measured using the IRAF task
{\it noao.onedspec.splot}. The continuum of a few objects showed the
thermal shape typical for ellipticals. In these cases a local
continuum was fitted to the single lines. The $1\sigma$ uncertainties
on the equivalent width and line flux values associated with the
placement of the continuum are estimated to be typically $\sim 30$ per
cent and $\sim 20$ per cent respectively. The `small bump' (which
spans the $2000 - 4000$~\AA~rest frame wavelength region and is a
blend of \FeII~lines and Balmer continuum emission) was present in
only 11 of the objects with a clearly detected \OII~$\lambda 3727$
emission line. In these cases the \OII~equivalent width has been
determined by taking the line flux relative to the (upper) continuum
including this feature as well as to the (lower) one fitted to the
entire spectrum. For these objects the typical difference between the
two equivalent width values is $\sim 20$ per cent. Since this
difference is rather low compared to the measurement errors and the
feature is not present in most objects, for consistency reasons the
`uncorrected' values are listed for these sources as well. Owing to
the relatively modest S/N of our spectra we have not attempted to
quantify the iron contamination of the \OIII~$\lambda 5007$ emission
line. Gaussian line profiles have been assumed for the narrow emission
lines \OII~$\lambda 3727$ and \OIII~$\lambda 5007$. In the case of the
broad emission lines \Mg~$\lambda 2798$, \Hb~$\lambda 4861$, and
\Ha~$\lambda 6563$ we have not assumed a specific line profile but
have intergrated the line flux over the entire emission line
range. This range has been determined for each object individually
using the quasar composite of \citet{Fra91} as a guide.

We have derived $2\sigma$ upper limits on the rest frame equivalent
widths and line fluxes when the lines were not detected but their
position was covered by the spectrum. In the case of narrow emission
lines we have determined non-detection limits only if the resolution
of the spectrum was high enough to allow for a detection of an
emission line with $FWHM \le 1000$ km s$^{-1}$. The non-detection
limits have been calculated assuming a rectangular emission line. The
emission line width has been assumed to be 1000 km s$^{-1}$ for the
narrow lines \OII~and \OIII. In the case of the broad lines \Mg, \Hb,
and \Ha~we have assumed for FSRQ values 5800 km s$^{-1}$, 4000 km
s$^{-1}$, and 4500 km s$^{-1}$ respectively, and for BL Lacs values
4100 km s$^{-1}$, 1500 km s$^{-1}$, and 1500 km s$^{-1}$ respectively.
These correspond to the average values for the detections. We have not
attempted to derive upper limits if the S/N at the position of the
line was $\la 3$.

The spectra of a number of DXRBS sources have not been observed at
parallactic angle (but typically close to the meridian). We have
chosen to include in our analysis line measurements based on these
spectra only if the flux loss due to atmospheric differential
refraction at the position of the line computed following
\citet{Fil82} was less than 30 per cent.

The narrow line region (NLR) luminosity for our sources has been
calculated following \citet{Raw91} as

\begin{equation}
L_{\rm NLR} = 3 \times (3 \times L_{\rm \OII} + 1.5 \times L_{\rm \OIII}),
\end{equation}

\noindent
where $L_{\rm \OII}$ and $L_{\rm \OIII}$ are the line luminosities of
\OII~$\lambda 3727$ and \OIII~$\lambda 5007$ respectively. For objects
with redshifts $z<0.5$ for which the spectrum did not cover the
position of \OII, $L_{\rm \OII}$ has been calculated from the linear
correlation between the ratio $L_{\rm \OII}/L_{\rm \OIII}$ and $\log
L_{\rm \OIII}$ found by \citet{Sau89}. In all other cases the relation
$L_{\rm \OIII} = 4 \times L_{\rm \OII}$ has been used when the
spectrum covered the position of only one of the two emission
lines. The broad line region (BLR) luminosity was calculated following
\citet{Cel97} as:

\begin{equation}
L_{\rm BLR} = \sum_i L_{i,obs} \frac{\langle L^{\ast}_{\rm BLR}
\rangle}{\sum_i L^{\ast}_{i,est}},
\end{equation}

\noindent
where $\sum_i L_{i,obs}$ is the sum of the measured luminosities of
the observed broad lines, scaled by the ratio of the estimated total
broad line region luminosity $L^{\ast}_{\rm BLR}$ to the estimated
luminosities of the observed broad lines. Both estimates were taken
from the results of \citet{Fra91}, and in the case of \Ha~$\lambda
6563$ from \citet{Gas81}.

\section{The Limitations of the Current Classification Scheme} \label{scheme}

We have initially classified objects in the DXRBS sample as BL Lacs
and FSRQ following the scheme proposed by \citet{Marcha96}. These
authors suggested to use the strength of the Ca H\&K break and the
equivalent width of the strongest {\it observed} emission line to
separate classes of radio-loud AGN.

The Ca H\&K break is a prominent absorption feature typically seen in
the spectra of elliptical galaxies (the hosts of radio-loud AGN; e.g.,
Wurtz et al. 1996\nocite{Wur96}) and is located at $\sim$
4000~\AA~rest frame wavelength. It is defined as $C = (f_+ - f_-) /
f_+$, where $f_-$ and $f_+$ are the fluxes in the rest frame
wavelength regions $3750 - 3950$ \AA~and $4050 - 4250$
\AA~respectively. Its value in normal non-active ellipticals is on
average $\sim$ 0.5 \citep{Dre87}. In blazars the value of the Ca H\&K
break is assumed to be decreased by the presence of non-thermal jet
emission.

An increase of the non-thermal jet continuum, however, will also lead
to a decrease of the equivalent width of any blazar emission line
(assuming a roughly constant ionizing continuum and so line
flux). Therefore, March\~a et al. proposed to use the Ca H\&K break
value -- equivalent width plane to classify radio-loud AGN. In
particular, they proposed to classify in this plane objects with Ca
H\&K break values $C \le 0.4$ and $C>0.4$ as blazars and radio
galaxies respectively. This was based on their reinvestigation of the
sample of \citet{Dre87} which yielded that less than 5 per cent of
non-active ellipticals had Ca H\&K break values $C \le 0.4$. In a
previous paper \citep{L02} we have shown that the transition in Ca
H\&K break value from blazars to radio galaxies is rather continuous,
since this feature is decreased by the amplified jet due to a change
in viewing angle. However, our studies showed that low-luminosity
radio-loud AGN become increasingly radio core-dominated around $C \sim
0.35$, a value similar to the one proposed by March\~a et al. to
separate blazars from radio galaxies but with a more physical meaning.

The blazar class itself was divided by March\~a et al. using a
diagonal line in the Ca H\&K break value -- equivalent width plane
(see Fig. \ref{marcha}). This line represented the simulated increase
in emission line equivalent width with decreasing non-thermal jet
continuum (i.e., with increasing Ca H\&K break value) for the
well-known BL Lac object 3C 371. These authors assumed that 3C 371 was
representative for the BL Lac class and argued that all objects with
weaker emission lines (left of the diagonal line) should be classified
as BL Lacs, whereas those with stronger emission lines (right of the
diagonal line) should be classified as quasars.

Using the Ca H\&K break value -- equivalent width plane to classify
radio-loud AGN the way it was suggested by March\~a et al. is
problematic for the following reasons:

\begin{enumerate}

\item
The diagonal line that separates BL Lacs and FSRQ is not only
arbitrary, but might contribute to the observed rareness of
high-redshift BL Lacs in present samples (see below).

\item
If the Ca H\&K break value -- equivalent width plane is to be used to
separate both blazars and radio galaxies it has to be applied to
narrow emission lines (and not to the strongest {\it observed}~
emission line), since only these are common to both.

\item
In any case, the rest frame equivalent width of broad emission lines
is not necessarily expected to decrease with decreasing Ca H\&K break
value, if this is indeed a suitable viewing angle indicator
\citep{L02}.

\end{enumerate}

\noindent
In the following subsections we expand on these points.

\subsection{The Redshift Effect}

\begin{figure}
\centerline
{\psfig{figure=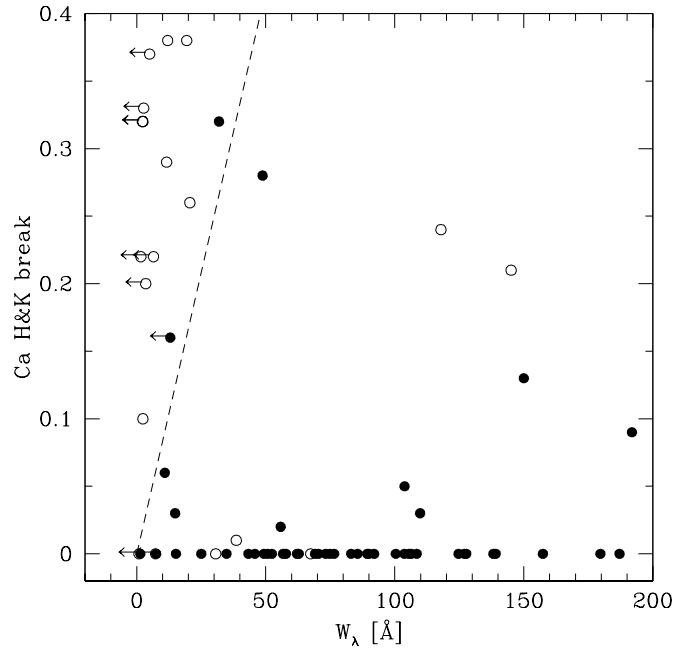,width=9cm}}
\caption{\label{marcha} The Ca H\&K break value versus the rest frame
equivalent width of the strongest observed emission line for DXRBS BL
Lacs and FSRQ. Open and filled circles indicate objects with redshifts
$z < 0.4$ and $z \ge 0.4$ respectively. Arrows indicate upper
limits. WGAJ0322.1-5205, WGAJ0539.0-3427, WGAJ0546.6-6415, and
WGAJ0631.9-5404 are off the plot to the right. The dashed line
represents the division between BL Lacs and FSRQ proposed by
\citet{Marcha96}.}
\end{figure}

In Fig. \ref{marcha} we plot the Ca H\&K break value versus the rest
frame equivalent width of the strongest observed emission line for the
objects in our sample. We have measured Ca H\&K break values in
spectra plotted as $f_\lambda$ versus $\lambda$ and have derived their
$1\sigma$ errors based on the S/N blueward and redward of the feature
(Tables \ref{bllacs} and \ref{fsrq}). We consider the Ca H\&K break to
have reached its minimum value of zero when the flux blueward of this
feature is equal to or larger than the one redward. The Ca H\&K break
is observable in the optical spectrum up to a redshift of $z \sim
1.2$. Therefore, we have included in Fig. \ref{marcha} all BL Lacs and
FSRQ from our sample with redshifts below this value, with the
exception of 16 objects (the spectrum of 10 objects does not cover the
Ca H\&K break location, and for 6 objects the Ca H\&K break is located
at the position of strong telluric absorption or is sampled with poor
S/N). Due to the redshift restriction the strongest emission lines
observed in these objects are \Mg~$\lambda 2798$, \OII~$\lambda 3727$,
\Hb~$\lambda 4861$, \OIII~$\lambda 5007$, and \Ha~$\lambda 6563$.

We first notice that the large majority (44/70 or 62.9 per cent) of
the sources plotted in Fig. \ref{marcha} have a Ca H\&K break diluted
to its minimum value of zero. This is not surprising, since
efficiently selected blazar samples are expected to include mainly
strongly beamed sources, i.e., sources with Ca H\&K break values of
$C=0$. At $C=0$, however, the diagonal line suggested by March\~a et
al. to separate blazars (dashed line in Fig. \ref{marcha}) does not
allow any {\it emission} lines for BL Lacs. Moreover, most sources
with $C=0$ are not expected to have detectable {\it absorption} lines,
since these will be diluted by the beamed jet emission. It then
follows that for efficiently selected blazar samples the current
classification scheme defines BL Lacs as those sources without {\it
  any} feature.

This has strong implications for the redshift distribution of current
blazar surveys, which also include weakly beamed sources (i.e.,
sources with $C > 0$). As discussed above, only weakly beamed BL Lacs
will have a redshift determination, while no restriction applies for
FSRQ. In flux-limited samples, however, strongly beamed sources will
have higher powers and so will be detected at higher redshifts.
Indeed, in Fig. \ref{marcha}, 48 objects have a redshift $z \ge 0.4$
(filled circles) and of these 38 objects (or 79.2 per cent) have a Ca
H\&K break value $C=0$, i.e., are highly beamed. As a consequence, the
March\~a et al. dividing line biases the BL Lac redshift distribution
towards lower values. A similar bias against high-redshift BL Lacs was
not introduced by the `original' classification criteria of the EMSS
and 1 Jy surveys. These allowed emission features (with $W_{\lambda} <
5$~\AA~) for sources with $C=0$.

In this context we would like to point out that, unlike DXRBS, other
recent blazar surveys did not strictly follow the March\~a et
al. classification scheme. For example, RGB \citep{LM99}, REX
\citep{Cac99} and CLASS \citep{Cac02} classified as BL Lacs all
sources with $C<0.25$ and $W_{\lambda} < 5$~\AA, i.e., they allowed
emission lines at $C=0$, contrary to what the March\~a et al. dividing
line suggests. Therefore a severe redshift bias due to classification
is not expected in these surveys.

\subsection{The Ca H\&K Break Value -- Equivalent Width Plane} \label{bimodal}

\begin{figure}
\centerline {\psfig{figure=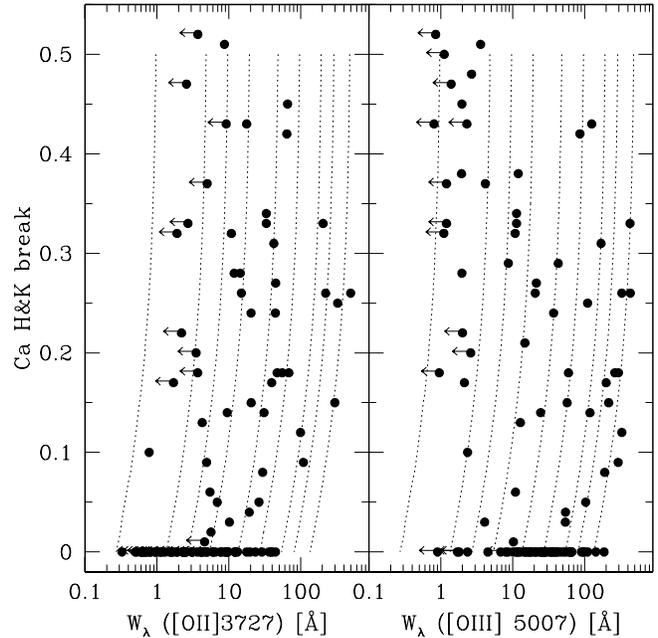,width=9cm}}
\caption{\label{dxrbs2jyoiioiii} The Ca H\&K break value versus the
\OII~$\lambda 3727$ (left panel) and \OIII~$\lambda 5007$ rest frame
equivalent width (right panel) for DXRBS and 2 Jy objects. Arrows
indicate upper limits. Dotted lines represent the simulated decrease
in rest frame equivalent width with increasing non-thermal continuum
contribution (see text for details) for a starting value (at $C=0.5$)
of W$_{\lambda} = 1$, 5, 10, 20, 50, 100, 200, 300, and 500~\AA~(from
left to right).}
\end{figure}

\begin{figure}
\centerline {\psfig{figure=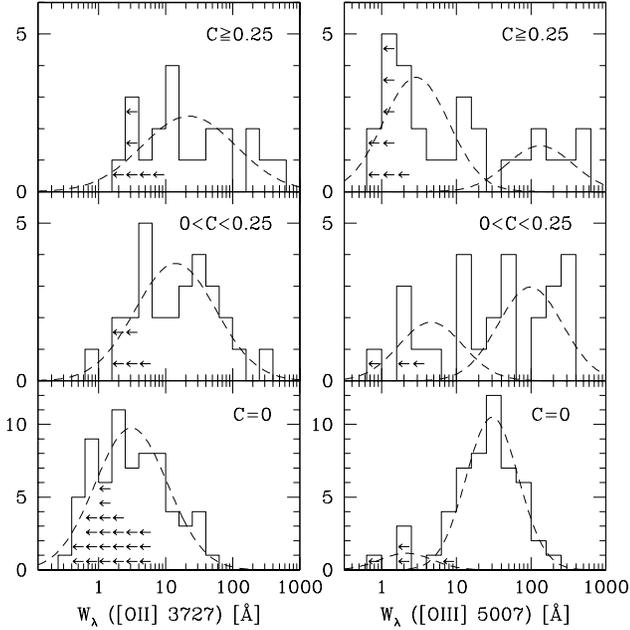,width=9cm}}
\caption{\label{oiioiiihisto}
  The \OII~$\lambda 3727$ and \OIII~$\lambda 5007$ rest frame
  equivalent width distributions for objects with $C=0$, $0 < C
  <0.25$, and $C \ge 0.25$ from the DXRBS and 2 Jy survey. Arrows
  indicate upper limits. The dashed curves represent the best-fit
  Gaussian models.}
\end{figure}

We now want to investigate if it is in general physically justified to
separate blazars in the Ca H\&K break value -- equivalent width plane.

Recently, we have shown that the Ca H\&K break value of BL Lacs and
low-luminosity radio galaxies decreases with viewing angle
\citep{L02}. This, however, means that only the equivalent widths of
narrow emission lines, believed to be isotropic, are expected to
decrease with Ca H\&K break value. The broad emission lines, on the
other hand, will be obscured by the putative circumnuclear dusty torus
at relatively large viewing angles, and will come progressively into
our line of sight as the viewing angle, and therefore the Ca H\&K
break, decreases. Simultaneoulsy we also expect the continuum flux to
increase (owing to the beamed jet component). This combination of
emission line and continuum flux increase renders the resulting
equivalent width and its dependence on the Ca H\&K break value
difficult to quantify and the possibility remains that the equivalent
widths of broad emission lines {\it increase} with Ca H\&K break
value. Moreover, if the Ca H\&K break value -- equivalent width plane
is to be used to separate also radio galaxies and blazars (as was
suggested by March\~a et al.) it is obvious that it has to be applied
to narrow emission lines only. Thus the question that we want to
answer becomes: Is it physically justified to separate blazars in the
Ca H\&K break value -- equivalent width plane using their {\it narrow}
emission lines? In the following we consider the strongest of the
narrow emission lines typically seen in radio-loud AGN, namely
\OII~$\lambda 3727$ and \OIII~$\lambda 5007$.

For our studies we have first simulated the decrease in equivalent
width with decreasing Ca H\&K break value. We have assumed at $C=0.5$
starting equivalent width values $W_{\lambda} = 1$, 5, 10, 20, 50,
100, 200, 300, and 500~\AA, and have increased the non-thermal jet
emission $f_{\rm jet}$ relative to a constant host galaxy flux $f_{\rm
gal}$. The corresponding Ca H\&K break value $C$ was calculated using
the relation $\log f_{\rm jet}/f_{\rm gal} = -3.74 \times C + 0.43$,
which results from the simulations presented in \citet{L02} for a jet
of optical spectral index $\alpha_{\nu} = 1$ (see their Fig. 2). The
resulting correlations between Ca H\&K break value and equivalent
width are represented by the dotted lines in
Fig. \ref{dxrbs2jyoiioiii}. We note that, contrary to the results of
March\~a et al., the above relation between Ca H\&K break value and
jet emission gives a non-linear dependence between the Ca H\&K break
and equivalent width values. For a change in the optical spectral
index of the jet of $\pm1$ the predicted equivalent widths at $C=0$
change by a factor of $\sim 2$.

Our simulations show that the equivalent width range of objects with
high Ca H\&K break values determines the maximum equivalent width
range of objects with low Ca H\&K break values. Note that it is the
maximum range, since an increasing jet emission can further decrease
the equivalent width, although the Ca H\&K break value has reached its
minimum value of zero. This means that the equivalent width range of
blazars depends on the (intrinsic) equivalent width range of their
parent radio galaxy populations. A separation of blazars into
subclasses using, e.g., one of the simulated dashed lines, would then
be physically justified only if there also existed two populations of
radio galaxies with significantly different \OII~and/or
\OIII~equivalent width distributions. In other words, we need to
observe a bimodal \OII~and/or \OIII~line luminosity distribution
intrinsic to the entire class of radio-loud AGN, which is expected to
manifest itself as a bimodal equivalent width distribution at any
given orientation. Note that equivalent width, although orientation
dependent in radio-loud AGN, is independent of redshift\footnote{This
is not strictly true for sources with continuua dominated by the host
galaxy, i.e., for sources with $C\ge0.25$ (see $\S$ \ref{classif}).}
and so more appropriate than line luminosity to quantify in a
meaningful way such a bimodality.

In order to investigate the existence of a bimodality intrinsic to the
class of radio-loud AGN we need a large number of sources that span a
wide range of equivalent widths and Ca H\&K break values. Therefore,
we have added to our sample SSRQ and radio galaxies from the DXRBS
(see Sections \ref{ssrqsec} and \ref{rgsec}). Given its radio spectral
index cut ($\alpha_{\rm r} \le 0.7$), however, DXRBS selects against
most radio galaxies. We have then included also the sample of radio
galaxies and quasars from the 2 Jy survey presented by \citet{Tad93}
and \citet{Morg97}. We believe that the use of objects from radio
surveys with different flux limits is warranted and should not
introduce any bias since line flux was not part of the selection
criteria.

In Fig. \ref{dxrbs2jyoiioiii} we plot the Ca H\&K break value versus
the rest frame equivalent width of \OII~$\lambda 3727$ (left panel)
and \OIII~$\lambda 5007$ (right panel) for radio-loud AGN from the
DXRBS and 2 Jy survey. In the left (right) panel of Fig.
\ref{dxrbs2jyoiioiii} we have included 77 (61) and 39 (45) sources
from the DXRBS and 2 Jy sample respectively. We note that we have
excluded from our studies sources with errors on the Ca H\&K break
value $> 0.2$. In Fig. \ref{oiioiiihisto} we show the equivalent width
distributions for these objects, binned into three groups of Ca H\&K
break values $C=0$, $0<C<0.25$ and $C \ge 0.25$. These groups comprise
67, 28 and 21 objects respectively in the case of \OII, and 52, 26 and
28 objects respectively in the case of \OIII.

In order to quantify a possible bimodality in these distributions we
have used the \kmm~algorithm \citep{Ash94}. The \kmm~algorithm
computes for a given univariate dataset the confidence level at which
the single Gaussian model can be rejected in favor of a two Gaussian
model. We have fitted homoscedastic groups (i.e., groups with similar
covariances), and only if the resulting confidence level was below 95
per cent have we also considered the heteroscedastic case. For the
distributions in Fig. \ref{oiioiiihisto} we get a high confidence
level for the rejection of the single Gaussian model in the case of
\OIII~for objects with $C \ge 0.25$ ($P=99.0$ per cent) and $C=0$
($P=98.5$ per cent), and a lower significance for objects with
$0<C<0.25$ ($P=90.2$ per cent). For \OII~a single Gaussian model is
always the best fit. We have overlaid the resulting best-fit Gaussians
as dashed lines in Fig. \ref{oiioiiihisto}. We note that the
\kmm~algorithm does not consider censoring in the data which is
present for both \OII~and \OIII~in all Ca H\&K break value bins. Since
these limits are {\it all} at the lower end of the distributions, the
significance of the bimodality would increase in all cases if these
limits were taken into account. Therefore, a bimodality cannot be
excluded also in the case of \OII.

The single Gaussian models for the \OII~equivalent width distributions
give mean values of about 22, 14, and 3~\AA~for objects with $C \ge
0.25$, $0<C<0.25$, and $C=0$ respectively. This observed decrease in
equivalent width is well reproduced by our simulations that give at
$C=0$ a value of $\sim 5$~\AA~if we assume a starting value of $\sim
20$~\AA~at $C=0.5$. This further confirms the use of the Ca H\&K break
as a statistical beaming indicator. In the case of \OIII~we get mean
values for the two best-fit Gaussians of $\sim 3$ and 135~\AA~for
objects with $C \ge 0.25$, and $\sim 5$ and 100~\AA~for objects with
$0<C<0.25$. For starting values of $\sim 5$ and 140~\AA~our
simulations predict values of $\sim 1.5$ and 40~\AA~at $C=0$. This is
similar to the means of $\sim 2$ and 30~\AA~observed for objects with
$C=0$.

Therefore, although the observed bimodality in the case of sources
with $0<C<0.25$ is only marginally significant (but see above), the
\OIII~bimodality for radio-loud AGN appears to have physical
reality. Relativistic beaming simulations give values in good
agreement with the observed decrease of the mean of the two modes with
Ca H\&K break. Assuming no bimodality for objects with
$C\ge0.25$ would result in a best-fit single Gaussian model with a mean of
$\sim 10$~\AA, which is {\it smaller} than the mean of $\sim
20$~\AA~that results for a best-fit single Gaussian model for objects
with $C=0$. In other words, only a bimodality for objects with $C \ge
0.25$ can account for the values observed in blazars within the
beaming scenario. We note, however, that objects with $C=0$ and very
small \OIII~equivalent widths seem to be underrepresented in our
sample. The missing sources are most likely the featureless BL Lacs
with no available redshift. In the DXRBS there are 13 such objects (5
of which have been observed by us) that have been excluded from our
analysis. \citet{Tad93} failed to identify recognizable features for
11 sources out of their complete sample of 87 2 Jy sources.

\section{A Physical Classification Scheme} \label{weakstrong}

\begin{figure*}
\begin{minipage}{80mm}
\centerline
{\psfig{figure=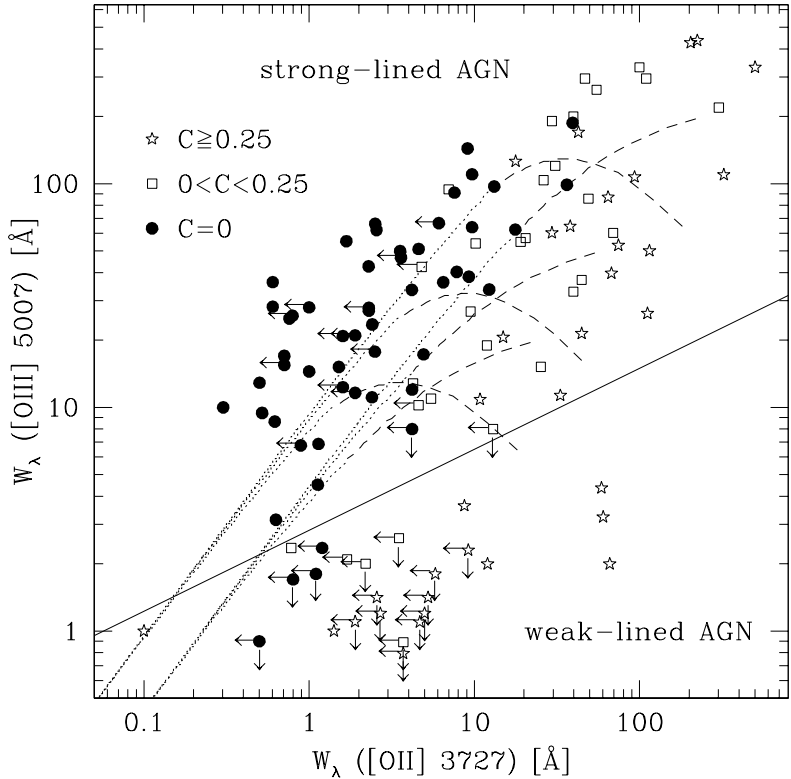,width=11cm}}
\caption{\label{beamingpart1} The rest frame equivalent widths of
\OIII~$\lambda 5007$ versus \OII~ $\lambda 3727$ for objects from the
DXRBS and 2 Jy survey. Filled circles, open squares and stars
represent sources with Ca H\&K break values $C=0$, $0<C<0.25$ and
$C\ge0.25$ respectively. Arrows indicate upper limits. The dashed and
dotted lines represent simulated loci of objects with $C>0$ and $C=0$
respectively. The solid line separates weak- (below the line) and
strong-lined radio-loud AGN (above the line) as defined in our studies
(Section \ref{classes}). See text for details.}
\end{minipage}
\end{figure*}

We want now to investigate the most appropriate way to separate all
radio-loud AGN (blazars and radio galaxies) into sources with {\it
  intrinsically} weak and strong \OIII~emission lines, which we dub
weak- and strong-lined radio-loud AGN respectively. In a future paper
(Landt et al., in prep.) we will investigate in detail the possible
underlying physical differences between the two classes.

\subsection{The \OIII~--~\OII~Equivalent Width Plane} \label{sep}

One option to split a sample of radio-loud AGN into weak- and
strong-lined sources would be to use as a dividing line one of the
simulated lines in the Ca H\&K break value -- \OIII~equivalent width
plot. Then, one could think of using simply the simulated line
corresponding at $C=0.5$ to the value of the intersection point of the
two best-fit Gaussians for objects with $C \ge 0.25$. However, a large
fraction (52/106 or 49 per cent) of the sources included in
Fig. \ref{dxrbs2jyoiioiii} (right panel) have $C=0$. For these objects
the dividing value at $C=0$ resulting from our simulations could be
problematic, since, as mentioned earlier, the equivalent width will
continue to decrease with increasing jet emission, although the Ca
H\&K break has reached its minimum value. In other words, at $C=0$
extremely beamed strong-lined AGN can cross the dividing line and
invade the region of weak-lined AGN, or, put differently, at $C=0$ it
becomes impossible to distinguish between `intrinsically' weak-lined
sources and those appearing weak because they are strongly
beamed. This means that the Ca H\&K break value -- equivalent width
plane simply does not offer the dynamic range necessary for a
meaningful separation of {\it all} radio-loud AGN into weak- and
strong-lined sources.

Such a separation, however, has to be based in any case on the
\OIII~equivalent width (the \OIII~line luminosity is not an option,
since it depends strongly on redshift), but this requires a method to
disentangle orientation effects if radio galaxies and blazars are
included. We suggest that this is possible using an \OIII~$\lambda
5007$~--~\OII~$\lambda 3727$ equivalent width plot. Since the
equivalent widths of both these emission lines are expected to
decrease with viewing angle, in such a plot objects viewed at larger
and smaller angles are expected to be concentrated at higher and lower
\OII~and \OIII~equivalent width values respectively, and so to
populate distinct regions of the plane.

In Fig. \ref{beamingpart1} we have plotted the \OIII~$\lambda 5007$
versus the \OII~$\lambda 3727$ rest frame equivalent widths for
objects from the DXRBS and 2 Jy survey (118 objects). In order to
increase the number statistics we have added objects with errors on
their Ca H\&K break values $>0.2$ (17 sources; mean error 0.3), and
objects with no available Ca H\&K break measurement (8 sources). We
have plotted objects with Ca H\&K break values $C\ge0.25$, $0<C<0.25$
and $C=0$ as stars (32 objects), open squares (30 objects), and filled
circles (48 objects) respectively. In cases where a Ca H\&K break
measurement was not available, we have assumed that objects are viewed
at relatively small angles if broad emission lines were present in
their optical spectra (6 objects, grouped with sources with $C=0$ and
plotted as filled circles), and that they are viewed at relatively
large angles if only narrow lines had been observed (2 objects,
grouped with sources with $C\ge0.25$ and plotted as stars).

Fig. \ref{beamingpart1} shows that, as anticipated, sources with
relatively high and low Ca H\&K break values, i.e., weakly and
strongly beamed sources respectively, separate in this plane. Also, as
argued in Section 4, the majority of our blazars appear to be beamed
radio galaxies with strong \OIII~emission lines, since it is mostly
this type of radio galaxies that have \OIII~and \OII~equivalent widths
higher than those of blazars. (Note that these quantities are expected
to decrease with viewing angle only if radio galaxies and blazars are
part of the same population.)

We now want to assess the most appropriate separation scheme for weak-
and strong-lined radio-loud AGN in the \OIII~--~\OII~equivalent width
plane applicable to {\it both} radio galaxies and blazars. A striking
feature of Fig. \ref{beamingpart1} is the clear-cut upper envelope
that extends from sources with high Ca H\&K break values to those with
low Ca H\&K break values, i.e., from radio galaxies to blazars, for
which we get roughly a slope of $\sim 0.4$. We assume that this
represents the general relation between the \OIII~and \OII~equivalent
widths of radio-loud AGN as the viewing angle changes, and have
attempted to reproduce it. We have assumed two components for the
continuum, jet and host galaxy, and let the jet emission increase
relatively to the host galaxy (jet/galaxy ratio defined at 5500~\AA)
as the angle with respect to the line of sight decreases. We have
assumed starting \OIII~equivalent width values $W_{\rm [OIII]} = 200$,
50 and 20~\AA~and a constant line luminosity ratio $L_{\rm
[OII]}/L_{\rm [OIII]} = 0.3$ (the mean value for the 84 objects in our
sample with both a detected \OII~and \OIII~emission line). For the jet
we have assumed a power-law spectrum with spectral index $\alpha_{\nu}
= 1$. Our results are shown in Fig. \ref{beamingpart1} as the
rightmost set of dashed and dotted lines, which indicate loci of
objects with Ca H\&K break values $C>0$ and $C=0$ respectively. For a
change in the optical spectral index of the jet of $\pm 1$ the
predicted equivalent widths for sources with $C=0$ change by a factor
of $\sim 2$.

Our simulations reproduce well the equivalent width decrease for
objects with $C>0$ as well as the position at which sources start to
have a Ca H\&K break value of zero. However, the predicted relation
between the \OII~and \OIII~equivalent widths in the regime of sources
with $C=0$ is steeper than the envelope to the data. In our
simulations the Ca H\&K break is diluted to its minimum value of zero
when the jet emission dominates at all optical wavelengths. As the jet
emission increases further both the \OII~and \OIII~equivalent widths
decrease in a similar way, resulting in a correlation between the two
with a slope of one. This is shown by the dotted lines in
Fig. \ref{beamingpart1} which converge irrespective of the starting
\OIII~equivalent width values. But the envelope assumed to represent
the general relation between the \OIII~and \OII~equivalent widths of
radio-loud AGN has a slope flatter than one. This means that with
decreasing viewing angle either the line luminosity ratio $L_{\rm
[OII]}/L_{\rm [OIII]}$ decreases or the continuum at the position of
\OII~increases more than the one at the position of \OIII.

The first possibility, namely that the line luminosity ratio $L_{\rm
[OII]}/L_{\rm [OIII]}$ decreases with viewing angle, could be the case
if, e.g., the \OIII~emission was anisotropic \citep{Hes96}. Indeed,
for the objects in our sample with a detection of at least one of the
two emission lines and errors on the Ca H\&K break value $<0.2$ we
find a strong ($P>99.9$ per cent) correlation between $L_{\rm
[OII]}/L_{\rm [OIII]}$ and Ca H\&K break value, albeit with a large
scatter. We have used here and in the following the ASURV analysis
package \citep{Iso86} whenever censoring was present in our data.

In order to test if an increase of the \OIII~line flux with decreasing
viewing angle can in fact account for the distribution of the points
in Fig. \ref{beamingpart1} we have repeated the simulations. We have
now assumed starting \OII~(and not \OIII~as before) equivalent widths
$W_{\rm \OII} = 200, 50$ and 20~\AA~and increased the \OIII~line flux
in dependence of the Ca H\&K break value using the above correlation
between $L_{\rm [OII]}/L_{\rm [OIII]}$ and Ca H\&K break value. The
results are shown in Fig. \ref{beamingpart1} as the leftmost set of
dashed and dotted lines, indicating loci of objects with Ca H\&K break
values $C>0$ and $C=0$ respectively. This new set of simulations seems
to reproduce the equivalent width decrease for objects with $C>0$
worse than the earlier one. Moreover, it still cannot reproduce the
upper envelope to the data. This is because the increase in jet
emission `catches up' with the \OIII~line flux increase, causing the
\OII~and \OIII~equivalent widths to decrease in a similar way as soon
as the Ca H\&K break is diluted to its minimum value.

The second possibility, namely that the continuum at the position of
\OII~increases with decreasing viewing angle more than the one at the
position of \OIII, already applies to sources with $C>0$. The
difference in spectral shape between the two continuum components, jet
and host galaxy, is such that in these sources the total continuum
increases mainly at shorter wavelengths. For sources with $C=0$,
however, where the jet emission dominates, this is only possible if we
assume a third component that starts to appear at UV frequencies at
the smallest viewing angles.

\begin{figure}
\centerline
{\psfig{figure=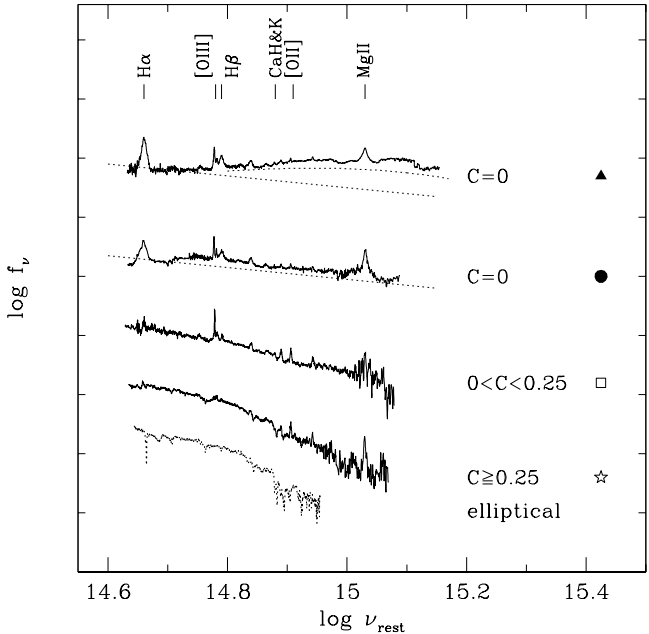,width=9cm}}
\caption{\label{composites} Composite spectra for objects from the
DXRBS and 2 Jy survey grouped according to their Ca H\&K break
value. Also shown is the spectrum of an elliptical galaxy. A power law
continuum with slope $\alpha_{\nu} = 1$ and a Balmer continuum are
underlaid as dotted lines to the composites for objects with Ca H\&K
break value $C=0$. Objects included in the composites are plotted in
Fig. \ref{beamingpart2} with symbols shown on the right side. Objects
with $C=0$ are separated into sources with optical spectral slopes
between the rest frame frequencies of \OII~and \OIII~below (top,
triangle) and above (bottom, circle) $\alpha_{\rm oiii}^{\rm oii} =
1$.}
\end{figure}

\begin{figure}
\centerline
{\psfig{figure=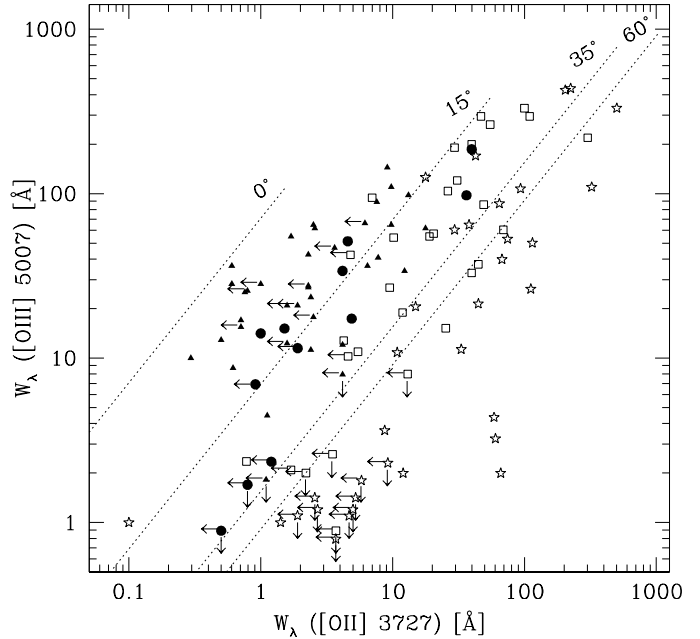,width=9cm}}
\caption{\label{beamingpart2} The rest frame equivalent widths of
\OIII~$\lambda 5007$ versus \OII~$\lambda 3727$ for objects from the
DXRBS and 2 Jy survey. Symbols are as in Fig. \ref{composites}. Arrows
indicate upper limits. Dotted lines represent loci of constant viewing
angle as labeled (see text for details).}
\end{figure}

The accretion disc believed to surround the central black holes of all
AGN, although neglected in our simulations, cannot be this additional
component. Assuming a constant emission line ratio, a linear relation
with a slope less than one between the logarithmic \OII~and
\OIII~equivalent width values implies that the continuum spectral
slope between the rest frame frequencies of \OII~and
\OIII~($\alpha_{\rm oiii}^{\rm oii}$) is not constant but decreases
with equivalent width. Then, for example, for a constant emission line
ratio of $L_{\rm [OII]}/L_{\rm [OIII]} = 0.3$ a slope of $\sim 0.4$
for the linear relation between the logarithmic \OII~and
\OIII~equivalent width values implies extremely `blue' optical
spectral slopes for objects with the smallest equivalent widths (e.g.,
$\alpha_{\rm oiii}^{\rm oii} \sim -4$ if $W_{\rm \OII} \sim 3$). This
means that the additional component needs either to have $\alpha_{\rm
  oiii}^{\rm oii}$ much flatter than one, i.e., much flatter than the
spectral slope assumed for the jet, or to cover only the location of
\OII. The accretion disc spectrum, however, is believed to extend over
a large range of frequencies covering the locations of both \OIII~and
\OII~and is in general approximated in the optical/near-UV regime by a
power law of the form $S_\nu \propto \nu^{1/3}$
\citep[e.g.,][]{Czerny87, Kor99}. Furthermore, \citet{Delia03} have
shown that the thermal (disc) component in DXRBS FSRQ makes up, on
average, only $\sim 15$ per cent of the optical/UV emission in these
sources.

In the following we want to assess if an additional component with the
required properties is indeed present in the spectra of radio-loud AGN
viewed at the smallest angles.

\subsection{The Continuum Emission of Radio-Loud AGN} \label{cont}

For our sources from the DXRBS and 2 Jy survey we have built composite
spectra for objects with Ca H\&K break values $C\ge0.25$, $0<C<0.25$
and $C=0$, the latter separated also into sources with $\alpha_{\rm
  oiii}^{\rm oii} \ge1$ and $<1$ (calculated from the ratios between
their \OII~and \OIII~continuum luminosities). Before combining them we
have shifted the spectra to their rest frame wavelength, normalized at
5500~\AA~and trimmed off noisy edges. We have merged the spectra
building an average with the IRAF task {\it noao.onedspec.scombine}.
The resulting four composites are displayed in Fig. \ref{composites}.
From top to bottom these contain 37, 12, 28 and 34 objects.

Our composites show that as expected the continuum of radio-loud AGN
flattens with decreasing Ca H\&K break value. The simplest
interpretation seems to be indeed that the observed flattening results
from different components dominating the continuum emission at
different orientations. A comparison between the composite for objects
with $C\ge0.25$ and the spectrum of an elliptical galaxy (dotted
spectrum in Fig. \ref{composites}) confirms that the galaxy component
dominates at large viewing angles. As the viewing angle decreases and
relativistic beaming enhances the jet emission, the continuum flattens
(composite for objects with $0<C<0.25$) until the jet emission
dominates (lower composite for objects with $C=0$ to which a power-law
spectrum with slope $\alpha_\nu = 1$ is overlaid as dotted
line). Finally, a group of sources with $C=0$ whose continuum emission
at UV frequencies is dominated by a third component is also present
(upper composite for objects with $C=0$). According to our earlier
considerations these sources are expected to be viewed at the smallest
angles.

The additional component is most likely the ultraviolet excess often
observed in AGN and interpreted as Balmer continuum/\FeII~emission
\citep[e.g.,][]{Mal82}. Following \citet{Wills85} we have then
calculated the Balmer continuum for a temperature of 15,000 K and have
overlaid the resulting spectrum as the dotted curve to the upper
composite for objects with $C=0$. A remarkable agreement is
evident. But is it feasible to assume that the Balmer continuum
emission starts to dominate at the smallest viewing angles?
Observations indicate that the Balmer continuum is located within the
BLR \citep{Maoz93}. Therefore, as the broad emission lines themselves
we expect the Balmer continuum to be obscured by the putative
circumnuclear dusty torus at large viewing angles and to come into our
line of sight only if the AGN is oriented at relatively small viewing
angles.

To explain the distribution of points in Fig. \ref{beamingpart1} we
have then repeated our simulations assuming now three components for
the continuum emission of radio-loud AGN, namely host galaxy, jet and
Balmer continuum, and increased the flux of the latter two with
decreasing viewing angle. We have simulated the change in orientation
by increasing the jet flux relative to the host galaxy emission
(jet/galaxy ratio defined at 5500~\AA) and calculated from this
viewing angles using our results presented in \citet{L02}. In
particular we have derived a relation between jet/galaxy ratio and
viewing angle using from these studies the linear relation between
jet/galaxy ratio and Ca H\&K break value for a jet with spectral slope
$\alpha_\nu = 1$, and the linear relation between Ca H\&K break value
and viewing angle for a jet with Lorentz factor $\Gamma=3$. We have
assumed that these relations are valid for all radio-loud AGN. The
flux increase of the Balmer continuum with decreasing viewing angle
has been adjusted interactively to give a slope of $\sim 0.4$ for the
correlation between the logarithmic \OIII~and \OII~equivalent widths.
The resulting loci of constant viewing angle are shown as dotted lines
in Fig. \ref{beamingpart2}. These have a slope of one, since we have
assumed for simplicity a constant line luminosity ratio $L_{\rm
\OII}/L_{\rm \OIII}$ which does not change with viewing angle, but the
envelope of these lines follows the slope of the envelope to the data
points.

\begin{figure}
\centerline
{\psfig{figure=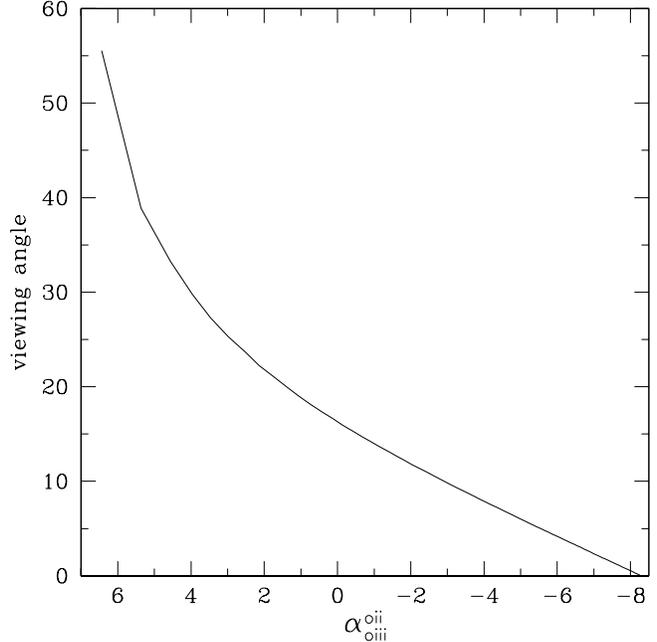,width=9cm}}
\caption{\label{VAcontratio} The relation viewing angle versus
continuum optical spectral slope (between the rest frame frequencies
of \OII~$\lambda 3727$ and \OIII~$\lambda 5007$) as obtained from our
simulations.
}
\end{figure}

These revised simulations predict that the continuum flux at the
position of \OII~is dominated by emission from the host galaxy, jet,
and Balmer continuum for objects viewed at angles $\phi \ga
35^{\circ}$, $ 15^{\circ} \la \phi \la 35^{\circ}$ and $\phi \la
15^{\circ}$ respectively. In Fig. \ref{beamingpart2} we have also
plotted objects included in the four composites with different
symbols. From this we see in particular that objects with $C=0$
included in the upper and lower composites (filled triangles and
circles respectively) cluster around the predicted values (and so are
indeed sources viewed at smaller and larger angles
respectively). Therefore, the continuum of radio-loud AGN appears to
change with viewing angle mainly at the position of \OII~rendering the
optical spectral slope $\alpha_{\rm oiii}^{\rm oii}$ a suitable
orientation indicator. Fig. \ref{VAcontratio} shows its relation to
viewing angle as obtained from our simulations. Note that $\alpha_{\rm
oiii}^{\rm oii}$ has more extreme values than those of the `usual'
optical continuum spectral slopes since it is measured over a much
narrower frequency range. The parameter $\alpha_{\rm oiii}^{\rm oii}$
is directly related to the Ca H\&K break value, the \OII~and
\OIII~emission lines being located blue- and redward of this feature
respectively, but it offers a much higher dynamical range.

In this context, we note that it has been argued that some radio-loud
AGN with weak emission lines might not possess significant broad line
regions and/or nuclear absorbing structures \citep[e.g.,][]{Chi99,
Chi02}. For these sources the relation between the parameter
$\alpha_{\rm oiii}^{\rm oii}$ and viewing angle shown in
Fig. \ref{VAcontratio} will not be valid. Moreover, in their case
$\alpha_{\rm oiii}^{\rm oii}$ will not be a 'better' orientation
indicator than the Ca H\&K break value, since for $C=0$ its value will
be simply the optical spectral index of the jet. Similarly, these
sources are expected to cluster around a line with slope of one (and
not of slope $\sim 0.4$) in the \OIII~--~\OII~equivalent width plane,
once $C=0$ is reached.

\subsection{Weak- and Strong-Lined Radio-Loud AGN} \label{classes}

We now want to use the \OIII~--~\OII~equivalent width plane to
separate our sources into weak- and strong-lined radio-loud AGN. For
the objects plotted in Fig. \ref{beamingpart1} we get \OIII~and
\OII~equivalent width distributions similar to the ones shown in Fig.
\ref{oiioiiihisto}. The intersection point of the two best-fit
Gaussians for objects with $C \ge 0.25$ is $W_{\rm \OIII} \sim
10$~\AA. We then define the line dividing weak- and strong-lined
radio-loud AGN in this plane to have a slope of 0.4 (assumed to
represent the general relation between the \OII~and \OIII~equivalent
widths of radio-loud AGN as the viewing angle changes) and to separate
objects with $C \ge 0.25$ and \OIII~equivalent width values below and
above 10~\AA~(solid line in Fig. \ref{beamingpart1}). This results in
a total of 25 and 93 objects in our sample being defined as weak- and
strong-lined radio-loud AGN respectively.

Of the weak- and strong-lined AGN 12 (13) and 56 (37) sources
respectively are part of the DXRBS (2 Jy survey). Based on DXRBS, BL
Lacs (as defined by Marcha et al.) are made up of $\sim 69$ per cent
(9/13) weak- and $\sim 31$ per cent (4/13) strong-lined radio-loud
AGN. The class of radio quasars, on the other hand, contains mostly
($\sim 98$ per cent or 46/47) strong-lined radio-loud AGN. Put
differently, weak-lined radio-loud AGN include $\sim 75$ per cent
(9/12) BL Lacs, $\sim 8$ per cent (1/12) quasars, and $\sim 17$ per
cent (2/12) radio galaxies, whereas strong-lined radio-loud AGN
include $\sim 82$ per cent (46/56) quasars, $\sim 7$ per cent (4/56)
BL Lacs, and $\sim 11$ per cent (6/56) radio galaxies. However,
because of its radio spectral index cut ($\alpha_{\rm r} \le 0.7$),
DXRBS selects against most radio galaxies. Therefore, considering only
strongly beamed sources, i.e., sources with Ca H\&K break values
$C=0$, $75^{+134}_{-49}$ per cent (3/4; errors are $1 \sigma$
uncertainties based on Poisson statistics) of the ones within the
weak-lined class are BL Lacs, and $97^{+29}_{-22}$ per cent (36/37) of
the ones within the strong-lined class are quasars. This means that
the overall properties of beamed weak- and strong-lined radio-loud AGN
are expected to be similar to those of the previously defined BL Lac
and quasar classes respectively, although no strong conclusion is
possible at this time for the weak-lined sources given the small number
statistics.

\section{Discussion}

\subsection{The Orientation Indicator $\alpha_{\rm oiii}^{\rm oii}$}

\begin{figure}
\centerline
{\psfig{figure=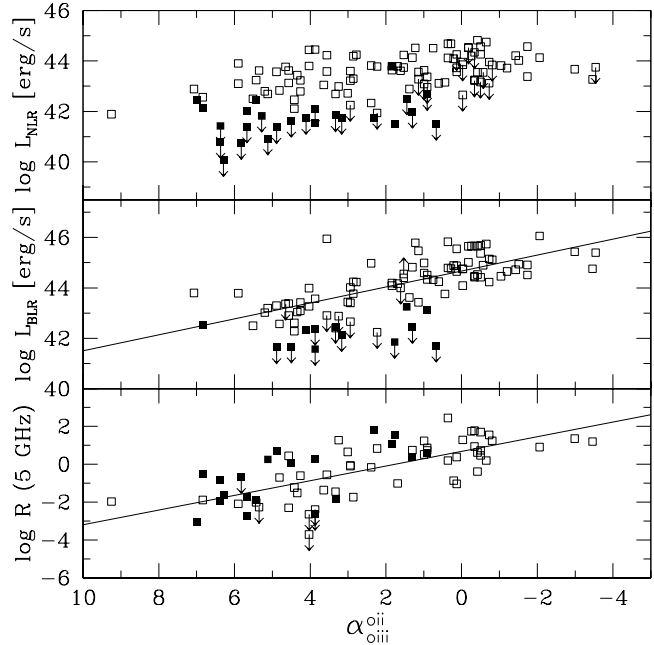,width=9cm}}
\caption{\label{contratio} The narrow line (upper panel) and broad
line region (middle panel) luminosities, and radio core dominance
parameter at 5 GHz (lower panel) versus the continuum optical spectral
slope (between the rest frame frequencies of \OII~$\lambda 3727$ and
\OIII~$\lambda 5007$) for objects from the DXRBS and 2 Jy
survey. Filled and open squares denote weak- and strong-lined
radio-loud AGN (as defined in our studies). Arrows indicate upper
limits. The solid lines represent the observed correlations for
strong-lined radio-loud AGN (middle panel) and for all sources (lower
panel).}
\end{figure}

Our interpretation that the \OIII~--~\OII~equivalent width plane is
suitable to disentangle orientation effects from intrinsic variations
in emission line studies of radio-loud AGN was based on the assumption
that the continuum optical spectral slope between the rest frame
frequencies of \OII~and \OIII~is indeed an orientation indicator. This
can be tested using, e.g., our estimates of the narrow line region
(NLR) and broad line region (BLR) luminosities, as well as
measurements of the radio core dominance parameter. The BLR of
radio-loud AGN, contrary to their NLR, is located much closer to the
central black hole, and so can be (partially or fully) obscured by the
thick, dusty torus believed to surround the nucleus in objects viewed
at large angles. This makes the BLR luminosity, unlike the NLR
luminosity, aspect dependent. Then, if $\alpha_{\rm oiii}^{\rm oii}$
is indeed an orientation indicator we expect it to correlate with the
{\it anisotropic} BLR but not with the {\it isotropic} NLR luminosity.
Similarly, we expect $\alpha_{\rm oiii}^{\rm oii}$ to correlate with
the radio core dominance parameter $R$ (defined as the ratio between
the core and extended radio flux), which is assumed to be a suitable
indicator of orientation for radio-loud AGN.

In Fig. \ref{contratio} we plot the BLR (middle panel) and NLR
luminosities (upper panel) versus $\alpha_{\rm oiii}^{\rm oii}$ for
our sample of weak- (filled squares) and strong-lined radio-loud AGN
(open squares). We have detected broad emission lines only for objects
with Ca H\&K break values $C \la 0.35$ (with the exception of 2Jy
1514$+$07, which has a narrow H$\alpha$ emission line with broad
wings; however, its Ca break has a large error, $C=0.6 \pm 0.5$).
Therefore, only these are included in Fig. \ref{contratio}, middle
panel, and we have assumed that this is the maximum Ca H\&K break
value (i.e., the maximum viewing angle) for which we can expect to
observe broad emission lines in the object's spectrum. (Note that this
value turns out to be the same as the one inferred by \citet{L02} to
be appropriate to separate core- and lobe-dominated low-luminosity
radio-loud AGN). We have then derived non-detection limits on the BLR
luminosity (as described in Section 3) only for objects with $C \le
0.35$. The NLR luminosity has been considered an upper limit if both
the \OII~and \OIII~emission lines were absent (17 objects) and also if
only the \OII~emission line could not be detected (16 objects).

Indeed, for our sample of strong-lined radio-loud AGN we find a strong
($P>99.9$ per cent) correlation between the BLR luminosity and
$\alpha_{\rm oiii}^{\rm oii}$ (solid line in Fig. \ref{contratio},
middle panel), which remains significant even if the common redshift
dependence is excluded. On the other hand, no significant ($P=91.1$
per cent) correlation is present between the NLR luminosity and
$\alpha_{\rm oiii}^{\rm oii}$ once the common redshift dependence is
excluded. We have used here the algorithms for partial correlation
analysis for censored data developed by \citet{Akr96}. Note that a
redshift dependence of $\alpha_{\rm oiii}^{\rm oii}$ is expected if
this is indeed a beaming indicator.\footnote{In a radio flux-limited
sample sources seen at smaller angles will be more strongly beamed and
will then have higher powers, which will result in them being detected
at redshifts higher than sources seen at larger angles.} Owing to the
large number of upper limits a similar analysis is not possible for
our sample of weak-lined AGN.

In Fig. \ref{contratio}, lower panel, we plot the radio core dominance
parameter at 5 GHz versus $\alpha_{\rm oiii}^{\rm oii}$ for part
(70/118) of our sample of weak- and strong-lined AGN. Radio core
dominance parameters for 2 Jy sources were taken from \citet{Morg97}
and for southern DXRBS sources calculated from our ATCA measurements
(Bignall et al., in prep.). We note that our ATCA observations were
conducted in snapshot mode which might have missed extended flux, and,
therefore, the DXRBS radio core dominance values are actually upper
limits to the true values. We have converted all values to a rest
frame frequency of 5 GHz assuming radio spectral indices of
$\alpha_{\rm r} = 0$ and 0.8 for the core and extended emission
respectively. The radio core dominance parameter strongly ($P>99.9\%$)
correlates with $\alpha_{\rm oiii}^{\rm oii}$, which further supports
its use as a statistical orientation indicator. In this respect, we
note that the composites shown in Fig. \ref{composites} are very
similar to the ones presented by \citet{Bak95a} for a sample of
quasars selected from the Molongolo survey and grouped by $R$. The
trends observed with increasing $R$ are similar to the ones observed
with decreasing Ca H\&K break value (i.e., decreasing $\alpha_{\rm
oiii}^{\rm oii}$), namely a flattening of the continuum emission,
markedly stronger appearance of the Balmer continuum, and a decrease
of the \OII~and \OIII~equivalent widths. This confirms in particular
our interpretation that the upper and lower composites for sources
with $C=0$ represent radio-loud AGN viewed at smaller and larger
angles with respect to the line of sight. These are similar to their
composites for sources with $R \ge 0.1$ and $R<0.1$ respectively.

\citet{Bak97}, however, concluded that the main cause for the
flattening of the continuum of quasars at small viewing angles was a
decreased obscuration by the putative circumnuclear dusty torus. In
other words, the larger the viewing angle, the more the continuum was
steepened by dust extinction. In support of this interpretation
\citet{Bak97} quoted the strong correlation observed between the
H$\alpha$/H$\beta$ emission line ratios and optical spectral
index. The amount of dust extinction predicted by this correlation
seemed to explain the observed decrease in \OII~and \OIII~equivalent
widths with $R$ and the observed range of optical spectral slopes. An
orientation-dependent obscuration as the primary cause for the change
of the optical continuum slopes of radio-loud quasars with viewing
angle, however, is ruled out by the fact that we observe this trend
with decreasing Ca H\&K break value. The large Ca H\&K break values
typical for the host galaxies of radio-loud AGN, which generally have
'red' spectra, are decreased at small viewing angles by emission from
the additonal components jet and Balmer continuum, which typically
have 'blue' spectra. Therefore, the observed change of the optical
spectral slopes of radio-loud AGN with viewing angle can be fully
explained by their (expected) main components dominating the continuum
at different orientations. On the other hand, although we believe that
dust extinction is not the dominant parameter governing the continuum
shape of radio-loud AGN, it might play a role in the observed Balmer
decrement decrease. In fact, we have measurements on both H$\alpha$
and H$\beta$ emission lines for 14 objects from the sample plotted in
Fig.  \ref{beamingpart1} and we find for these, similar to
\citet{Bak97}, a significant ($P=99.1$ per cent) anticorrelation
between the H$\alpha$/H$\beta$ emission line and $\alpha_{\rm
oiii}^{\rm oii}$.

\subsection{The Continuity Between Blazar Subclasses}

\begin{figure}
\centerline
{\psfig{figure=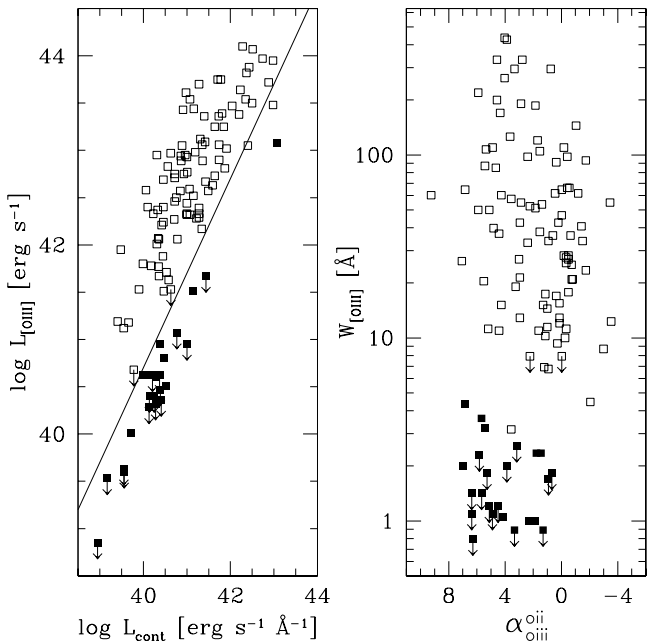,width=9cm}}
\caption{\label{SFplot} {\bf Left panel:} The \OIII~$\lambda 5007$
line luminosity versus the continuum luminosity underneath the line
for sources from the DXRBS and 2 Jy survey. Filled and open squares
indicate weak- and strong-lined radio-loud AGN respectively. Arrows
indicate upper limits. The solid line indicates a constant equivalent
width of 5~\AA. This plot is similar to Fig. 10 of Scarpa and Falomo
(1997). {\bf Right panel:} The \OIII~$\lambda 5007$ rest frame
equivalent width versus the continuum optical spectral slope (between
the rest frame frequencies of \OII~$\lambda 3727$ and \OIII~$\lambda
5007$) for objects from the DXRBS and 2 Jy survey. Symbols as in left
panel. Note that both panels include the same objects. See text for
details.}
\end{figure}

\citet{Sca97} investigated if BL Lacs and high-polarization quasars
(HPQ) had different intrinsic emission line luminosities considering
data for the broad emission line \Mg~$\lambda 2798$ for a sample of 34
sources. Their sample included 10 BL Lacs and 18 HPQ selected from the
literature, but newly observed by these authors, as well as 6 BL Lacs
from the 1 Jy sample of \citet{Sti93c}. Only sources with detected
\Mg~emission lines were considered. The approach Scarpa and Falomo
chose to address the problem was to plot the logarithmic line
luminosity versus continuum luminosity below the line (see their
Fig. 10). In such a plot lines of constant equivalent width form
diagonals with a slope of one. By plotting also the line of constant
equivalent width of 5~\AA~(the dividing value between BL Lacs and
quasars at the time) their plot showed that the transition between BL
Lacs and HPQ was continuous. Based on this, these authors argued that
from the point of view of emission line strengths it was not necessary
to invoke two different populations of blazars.

In Fig. \ref{SFplot}, left panel, we show a plot similar to that of
Scarpa and Falomo for our sample of weak- (filled squares) and
strong-lined radio-loud AGN (open squares) for the \OIII~$\lambda
5007$ emission line. The solid line indicates a constant equivalent
width of 5~\AA. In Fig. \ref{SFplot}, right panel, we have plotted for
the same objects the \OIII~rest frame equivalent width versus
$\alpha_{\rm oiii}^{\rm oii}$. These plots illustrate two important
points. First, {\it given a bimodal distribution for an emission
line}, which is the only physical justification for a separation of
radio-loud AGN based on emission line strength, the equivalent width
separating the two classes will depend on orientation. A constant
equivalent width value is not suitable for this purpose. In this
respect, Fig. \ref{SFplot}, right panel, illustrates clearly the
\OIII~bimodality (thus the need for two classes of radio-loud AGN) as
well as the dependence of the dividing equivalent width value on
orientation, of which $\alpha_{\rm oiii}^{\rm oii}$ is a suitable
statistical indicator. Second, a comparison between the plots in the
left and right panel of Fig. \ref{SFplot} shows that even when a
bimodality is present for an emission line of radio-loud AGN this will
not be clearly evident in a plot of line luminosity versus continuum
luminosity. Such a plot does not allow one to disentangle orientation
effects from intrinsic variations, a step which is, however, crucial
in studies involving emission lines of radio-loud AGN.

\subsection{The Classification of Radio-Loud AGN} \label{classif}

\begin{figure}
\centerline
{\psfig{figure=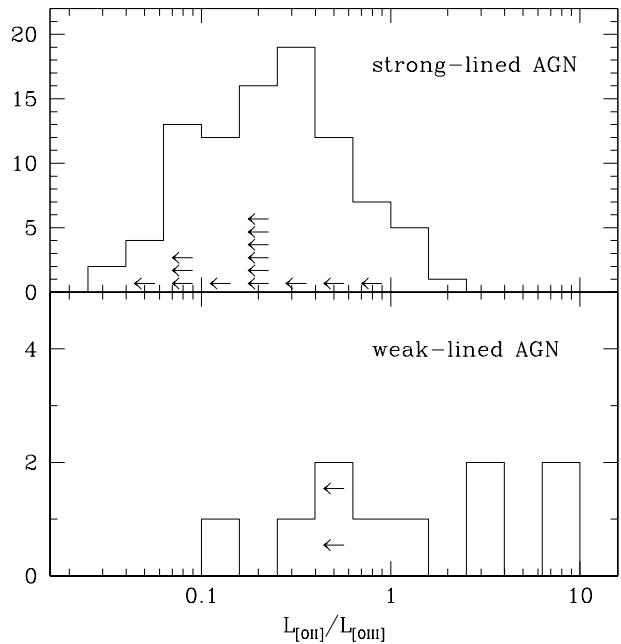,width=9cm}}
\caption{\label{lumratio} The distribution of the \OII~$\lambda 3727$
to \OIII~$\lambda 5007$ line luminosity ratio for weak- (lower panel)
and strong-lined radio-loud AGN (upper panel) from the DXRBS and 2 Jy
surveys. Arrows indicate upper limits on the \OII~line luminosity.}
\end{figure}

This work together with our earlier paper \citep{L02} represent a
complete physical revision of the current classification scheme for
blazars proposed by \citet{Marcha96}.

There have been previous attempts to separate radio-loud AGN, in
particular radio galaxies, based on their \OIII~emission line
strengths. However, in all these studies the separation value was
chosen rather arbitrarily. For example, \citet{Lai94} divided their
sample of FR II radio sources from the 3CR survey into low- and
high-excitation objects, which they defined to have \OIII~equivalent
widths below and above 3 \AA~respectively, with the latter required to
have also emission line ratios \OIII/\Ha~(narrow) $>0.2$. Similarly,
\citet{Jack97} classified radio galaxies in their subsample of 3CR and
3CRR radio sources as low- (LEG) and high-ionization narrow-line radio
galaxies (HEG) if they had \OIII~equivalent widths below and above 10
\AA~respectively. Radio galaxies with emission line ratios \OII/\OIII
$>1$ also fell in the former category. \citet{Tad98} distinguished in
their sample of radio sources from the 2 Jy survey between narrow-
(NLRG) and weak-line radio galaxies (WLRG), the latter defined as
sources with their spectra dominated by stellar absorption features
from the host galaxy and \OIII~emission lines weaker than 10~\AA. 

In this respect, we note that the separation value of $W_{\rm \OIII} =
10$~\AA~of \citet{Jack97} and \citet{Tad98}, although selected rather
arbitrarily, coincides with the intersection point of the two best-fit
Gaussians to the \OIII~equivalent width distribution of sources with
$C \ge 0.25$ plotted in Fig. \ref{beamingpart1} and so roughly with
our separation of weak- and strong-lined radio galaxies (see Section
\ref{classes}). However, our classification is not equivalent to the
separation of \citet{Jack97} into low- and high-ionization radio-loud
AGN. Fig. \ref{lumratio} plots the \OII/\OIII~emission line ratio
distributions for our weak- (lower panel) and strong-lined sources
(upper panel), excluding sources with upper limits on both their
\OII~and \OIII~emission line luminosities (10/25 and 91/93 objects
respectively). From this we see that, whereas most (85/91 or 93.4 per
cent) strong-lined AGN have ratios \OII/\OIII $<1$ and could be
classified as high-ionization sources, the class of weak-lined AGN
appears to be heterogeneous with equal number of sources having
\OII/\OIII $<1$ and $>1$.

Our new classification scheme, namely the \OIII~--~\OII~equivalent
width plane, is easily applicable to sources with $z \la 0.8$, for
which the optical spectrum (typically spanning the wavelength range
$3500 - 9000$~\AA) allows the detection of both the \OII~and
\OIII~emission lines. In this redshift range, actually, all one needs
in most cases is \OIII. Fig. \ref{SFplot}, in fact, shows that all
sources with $W_{\rm [OIII]} > 6$ \AA~are of the strong-lined type,
while all sources with $W_{\rm [OIII]} <3$ \AA~are of the weak-lined
type. Only when $W_{\rm [OIII]}$ is in between these two values one
needs \OII~as well. For sources at higher redshifts, however, infrared
spectra are required. In this respect note that even the
classification scheme proposed by \citet{Marcha96} requires infrared
spectra for sources with redshifts $z \ga 1.2$, for which the Ca H\&K
break is located outside the optical window.

Our new classification scheme is based on emission line equivalent
width and, therefore, in the case of radio galaxies (objects with
$C\ge0.25$) where the host galaxy dominates the continuum, it is
expected to depend also on spectroscopic constraints such as a final
slit width. A narrow slit will limit the fraction of galaxy light
included, and will lead to an overestimation of the equivalent widths
of these sources at low redshifts. This could introduce a bias against
low-redshift weak-lined AGN with $C\ge0.25$, since these will be
'misclassified' as strong-lined AGN. However, simulations show that
this kind of observational constraints are expected to affect our
classification scheme only for very nearby radio galaxies. Assuming an
elliptical host galaxy with a de Vaucouleurs surface brigthness
profile and a half-light radius of 10 kpc \citep[e.g.,][]{Pag03} we
get that for sources at redshifts $z\la0.07$ and $z\la0.03$ the
equivalent width will be overestimated by factors $\ga 5$ and $\ga 10$
respectively compared to the value measured at large redshifts. The
effects due to a difference in slit width are less severe. For
example, we expect the equivalent widths to be overestimated only by a
factor of $\sim 2$ if one uses a slit width of $1''$ instead of
$5''$.

\section{Summary and Conclusions}

We have used radio-loud AGN from the DXRBS and 2 Jy survey to
readdress the separation of blazars into BL Lacs and FSRQ as well as
the general classification of radio-loud AGN based on their emission
lines. In this respect, this work represents a physical revision of
the present classification scheme proposed by \citet{Marcha96}. Our
main results can be summarized as follows:

\begin{enumerate}

\item We have argued that a physically justified separation of blazars
  based on emission line strength requires the presence of a bimodal
  distribution for any of the {\it narrow} emission lines inherent to
  the entire class of radio-loud AGN. In our studies we have
  considered the equivalent width distributions of the narrow emission
  lines \OII~$\lambda 3727$ and \OIII~$\lambda 5007$ for $\sim 100$
  blazars and radio galaxies from the DXRBS and 2 Jy survey. We have
  found a bimodal distribution for the \OIII~emission line.
  
\item We have shown that all radio-loud AGN can be separated
  unambiguously into sources with {\it intrinsically} weak and strong
  \OIII~emission lines (dubbed weak- and strong-lined radio-loud AGN
  respectively) using an \OIII~--~\OII~equivalent width plane. This
  plane is suitable to disentangle orientation effects and intrinsic
  variations in radio-loud AGN. In most cases our classification
  scheme requires only \OIII, since all sources with $W_{\rm [OIII]} >
  6$ \AA~are of the strong-lined type, while all sources with $W_{\rm
    [OIII]} <3$ \AA~are of the weak-lined type.
  
\item We have presented composite spectra that illustrate that
  different components dominate the continuum emission of radio-loud
  AGN at different orientations. In particular, at shorter optical
  wavelengths the continuum emission is dominated by the host galaxy,
  jet emission and Balmer continuum for sources viewed at large,
  intermediate, and small angles respectively.
  
\item Based on DXRBS, the strongly beamed sources (i.e., sources with
  Ca H\&K break values of $C=0$) of the weak-lined class include $\sim
  75$ per cent BL Lacs, whereas those of the strong-lined class are
  mostly ($\sim 97$ per cent) quasars. This means that the overall
  properties of beamed weak- and strong-lined radio-loud AGN are
  expected to be similar to those of BL Lacs and quasars respectively.

\end{enumerate}

We note that our classification scheme is somewhat dependent on
observational constraints in the case of radio galaxies and weakly
beamed sources, due to their strong host galaxy component. However,
simulations show that this affects only relatively low redshift ($z
\la 0.1$) sources.

A bimodal equivalent width distribution for an emission line of
radio-loud AGN has not been reported so far. Therefore, this important
result needs further testing. This can be done, e.g., using the MRC/1
Jy radio sample, which contains a large number of sources ($\sim 550$
objects) with uniform spectroscopic observations, as well as current
on-going surveys, such as, e.g., RGB \citep{LM98, LM99}, REX
\citep{Cac99, Cac00}, and the low-redshift selected CLASS blazar
sample \citep{Marcha01, Cac02}. In a subsequent paper we plan to
investigate the physical differences between weak- and strong-lined
radio-loud AGN using their emission line measurements as well as
information on their radio, optical and X-ray luminosities.

\section*{Acknowledgements}
We thank Clive Tadhunter for providing the spectra of radio sources
from the 2 Jy survey in electronic format, and the anonymous referee
for suggestions which led to major improvements of our
manuscript. H.L. acknowledges financial support from the STScI DDRF
grants D0001.82260 and D0001.82299. E.P.  acknowledges support from
NASA grants NAG5-10109 and NAG5-9995, as well as NAG5-9997 (LTSA).

\small
\bibliography{/home/hermi/THESIS/thesis_references}

\end{document}